\documentclass[apj]{emulateapj}
\usepackage{natbib}
\shorttitle{The formation and disintegration of MBPs}
\shortauthors{Utz et al.}

\begin{document}


\title{THE FORMATION AND DISINTEGRATION OF MAGNETIC BRIGHT POINTS \\OBSERVED BY \textit{SUNRISE}/IMaX}


\author{D. Utz\altaffilmark{1}, J. C. del Toro Iniesta, L. R. Bellot Rubio}
\affil{Instituto de Astrof\'{i}sica de Andaluc\'{i}a (CSIC), Apdo. de Correos 3004, \\E-18080 Granada, Spain}
\email{utz@iaa.es, dominik.utz@uni-graz.at}
\altaffiltext{1}{Current address: IGAM-Kanzelh\"ohe Observatory, Institute of Physics, University
of Graz, AT-8010 Graz, Austria.}

\author{J. Jur\v{c}\'{a}k}
\affil{Astronomical Institute, Academy of Sciences of the Czech Republic, \\251 65 Ond\v{r}ejov, Czech Republic}


\author{V. Mart\'{\i}nez Pillet\altaffilmark{2}}
\affil{Instituto de Astrof\'{\i}sica de Canarias, V\'{\i}a L\'actea, s/n, E-38200 La Laguna, Spain}
\altaffiltext{2}{Current address: National Solar Observatory, Sunspot, NM 88349, USA.}

\author{S. K. Solanki\altaffilmark{3}}
\affil{Max-Planck Institut f{\"u}r Sonnensystemforschung, Max-Planck-Strasse, 2, D-37191, Germany}
\altaffiltext{3}{School of Space Research, Kyung Hee University, Yongin, Gyeonggi 446-701, Korea.}

\author{W. Schmidt}
\affil{Kiepenheuer-Institut f{\"u}r Sonnenphysik, Sch{\"o}neckstrasse 6, D-79104 Freiburg, Germany}





\begin{abstract}

The evolution of the physical parameters of magnetic bright points (MBPs) located in the quiet Sun (mainly in the interwork) during their lifetime is studied. 
First we concentrate on the detailed description of the magnetic field evolution of three MBPs. This reveals that individual features follow different, generally complex, and rather dynamic scenarios of evolution. Next we apply statistical methods on roughly 200 observed MBP evolutionary tracks. MBPs are found to be formed by the strengthening of an equipartition field patch, which initially exhibits a moderate downflow. During the evolution, strong downdrafts with an average velocity of 2.4 km/s set in. These flows, taken together with the concurrent strengthening of the field, suggest that we are witnessing the occurrence of convective collapses in these features, although only 30\% of them reach kG field strengths. This fraction might turn out to be larger when the new 4 m class solar telescopes are operational as observations of MBPs with current state of the art instrumentation could still be suffering from resolution limitations. Finally, when the bright point disappears (although the magnetic field often continues to exist) the magnetic field strength has dropped to the equipartition level and is generally somewhat weaker than at the beginning of the MBP's evolution. Also, only relatively weak downflows are found on average at this stage of the evolution. Only 16\% of the features display upflows at the time that the field weakens, or the MBP disappears. This speaks either for a very fast evolving dynamic process at the end of the lifetime, which could not be temporally resolved, or against strong upflows as the cause of the weakening of the field of these magnetic elements, as has been proposed based on simulation results. Noteworthy is that in about 10\% of the cases we observe in the vicinity of the downflows small-scale strong (exceeding 2 km/s) intergranular upflows related spatially and temporally to these downflows. The paper is complemented by a detailed discussion of aspects regarding the applied methods, the complementary literature and in depth analysis of parameters like magnetic field strength and velocity distributions. An important difference to magnetic elements and associated bright structures in active region plage is that most of the quiet Sun bright points display significant downflows over a large fraction of their lifetime (i.e., in more than 46\% of time instances/measurements they show downflows exceeding 1 km/s).
\end{abstract}


\section{Introduction}

Magnetic bright points (MBPs) are manifestations of small-scale solar magnetic fields \citep[for a recent review see, e.g.,][]{2009SSRv..144..275D} and are counted among the most interesting features on the Sun. Only isolated, small-scale magnetic field concentrations of kG strengths appear bright in the solar photosphere, as was shown by \citet{2014A&A...562L...1C} and \citet{2014A&A...568A..13R}. But the influence of the telescope PSF leads to a smoothing of the ``true'' magnetic field strength to lower values, so that recent observations display a distribution of field strengths from small fractions of kG up to 1.5 kG \citep[see, e.g.,][and references therein]{2007A&A...472..607B,2013A&A...554A..65U}. 
Such concentrations of magnetic field appear as bright point-like features in filtergrams and are termed MBPs \citep[e.g., ][]{2004A&A...422L..63W}. The increase in brightness can be explained due to a partial evacuation of the magnetic field structure leading to a decreased density within the structure. This decreased density turns into an effective lowering of the optical unity layer showing then deeper and hotter plasma. Besides, it allows more radiation and hence heat to penetrate into the feature from the surrounding hot walls \citep[see e.g.][]{2004ApJ...607L..59K,2004A&A...421..741V}.

MBPs have been studied for several decades with the first observations dating back to the 1970s \citep[see, e.g.,][]{1973SoPh...33..281D,1983SoPh...85..113M}. Since then their dynamical \citep[e.g.,][]{1996ApJ...463..365B,2005AA...441.1183D,2012ApJ...752...48C,2013A&A...549A.116J} as well as their statistical and magnetic properties \citep[e.g.,][]{2007A&A...472..607B,2008A&A...488.1101B,2013SoPh..tmp....7U,2013ApJ...770L..36G} have been characterised in increasing detail. In recent years such filtergram studies were done mostly in the Fraunhofer G-band \citep[][]{1998ApJ...506..439B,2004A&A...428..613B,2008A&A...488.1101B} but also in other molecular bands such as the CN band-head \citep[][]{2005A&A...437L..43Z} or the TiO \citep[][]{2010ApJ...725L.101A}. Besides the analysis of pure observational data also the comparison between characteristics in observations and simulations attract more and more interest \citep[e.g.,][]{2011ApJ...740L..40K}. Moreover, spectropolarimetric studies have contributed to our knowledge of MBPs \citep[e.g.,][]{2001ApJ...560.1010B,2008ApJ...677L.145N,2010ApJ...723..787V}. Lately, combinations of filtergram (e.g., G-band and white light) data and spectropolarimetric observations have been performed \citep[see, e.g.,][]{2010ApJ...723L.169R,2012SoPh..280..407R}.



The interest in MBPs arises because they are easy to observe proxies of kG magnetic flux concentrations. These in turn are of great interest for several reasons, such as the not yet completely understood chromospheric and coronal heating problem \citep[e.g.,][]{2003A&ARv..12....1W,2006SoPh..234...41K}. By connecting the solar interior to the outer solar atmosphere magnetic flux concentrations play an important role in most proposed heating mechanisms, irrespectively of whether the actual heating takes place via magnetic reconnection, ohmic dissipation, or wave heating processes. 
Another important aspect is the contribution of MBPs to variations in total solar irradiance, which may influence the Earth's climate \citep[see][for a review]{2013ARA&A..51..311S}. Furthermore, they influence the granulation itself and the energy transport. A number of authors including \citet[][]{1989ApJ...336..475T}, \citet{1989SoPh..119..229M}, \citet{2010A&A...524A...3N}, and \citet{2011ApJ...731...29A} have reported on smaller granular cells around ensembles of magnetic flux concentrations and concluded that the presence of strong magnetic fields suppresses to a certain extent normal convective flows. Furthermore there is recently a debate going on about a possible subpopulation of granules ---so-called mini granules \citep[e.g.,][]{2012ApJ...756L..27A,2014A&A...563A.107L}. Hence the question arises if and how this kind of granules is related to small-scale magnetic fields? Studies of the evolution of the number of network magnetic elements and MBPs in the quiet Sun over the solar cycle have given results that depend on the magnetic flux per feature and on the length of time considered \cite[see][]{2003ApJ...584.1107H}. Another recent attempt (for the last solar minimum) of tracking the number and variation of the number of individual MBPs was, unfortunately, not conclusive \cite[see][]{2011SoPh..274...87M}.

Since kG fields are needed to produce MBPs, their appearance, disappearance, and evolution teach us how the vigorous photospheric convection interacts with magnetic fields, strengthens them and keeps them in equilibrium until they dissolve or disintegrate. The most feasible theory for the creation of small-scale kG magnetic features is that of convective collapse proposed by \citet{1978ApJ...221..368P} or similar by \citet{1979SoPh...61..363S}. The convective collapse model is based on the idea that a concentrated magnetic field suppresses efficiently the transport of energy by convection. Therefore the plasma locally cools down and a downdraft sets in. This downdraft evacuates the flux tube, so that the excess pressure of the surrounding gas compresses it (or it ``collapses''). After the collapse, a new stable state can emerge (depending on the initial conditions) with magnetic field strengths above 1 kG. The convective instability acting in small-scale magnetic features and their subsequent concentration have been verified observationally \citep[e.g.,][]{2001ApJ...560.1010B,2008ApJ...677L.145N,2011A&A...529A..79N,2014ApJ...789....6R} as well as by numerical experiments \citep[see, e.g,][]{1998A&A...337..928G,2010A&A...509A..76D,2014A&A...565A..84H}.

While the convective collapse model serves us well in understanding the processes leading to kG flux tubes, much less is known about the processes leading to the dissolution and disintegration of the strong kG tubes. The aforementioned work of \cite{2001ApJ...560.1010B} states that disintegration might be caused by upward moving shocks in the flux tubes produced by fast downflowing material hitting the base of the flux tube and bouncing back. Such a mechanism was observed in simulations by \citet{1998ApJ...495..468S}. This scenario has been studied by computer simulations such as the one of \cite{1999ApJ...522..518T}. Another possibility for the disintegration of thin magnetic flux tubes is the interchange instability, which can be overcome by whirl flows around the flux tube \citep[see][]{1975SoPh...40..291P,1984A&A...140..453S}. Alternatively, the magnetic field may leak out of the features by diffusion, or may be torn out of them by turbulent convection, or may reconnect with turbulent fields in the immediate surroundings of the magnetic features. Hence investigating the evolution of the plasma parameters of magnetic flux concentrations during the very end of their lifetime is of great interest. How is the flux concentration maintained and stabilized during their life, and why and how does it dissolve?

To shed more light on these questions it is necessary to track individual magnetic flux concentrations during their lifetime and obtain as detailed and highly resolved information about the interesting quantities as possible. This is the main aim of the present study, where we employ MBPs as proxies of kG magnetic concentrations. These are used to distinguish such concentrations from other, weaker field, features. In the first part a few specific evolutionary tracks will be discussed in detail while later on a statistical approach is applied to all of the roughly 200 tracked MBPs. This work complements the recent work of \citet{2014ApJ...789....6R} in which the evolution of a single magnetic patch was studied in full detail and in addition the study of \citet{2014A&A...565A..84H}, who investigated in detail the formation, evolution, and dissolution of MBP features in MHD simulations.

\begin{figure*}

\begin{center}
\includegraphics[width=0.95\textwidth]{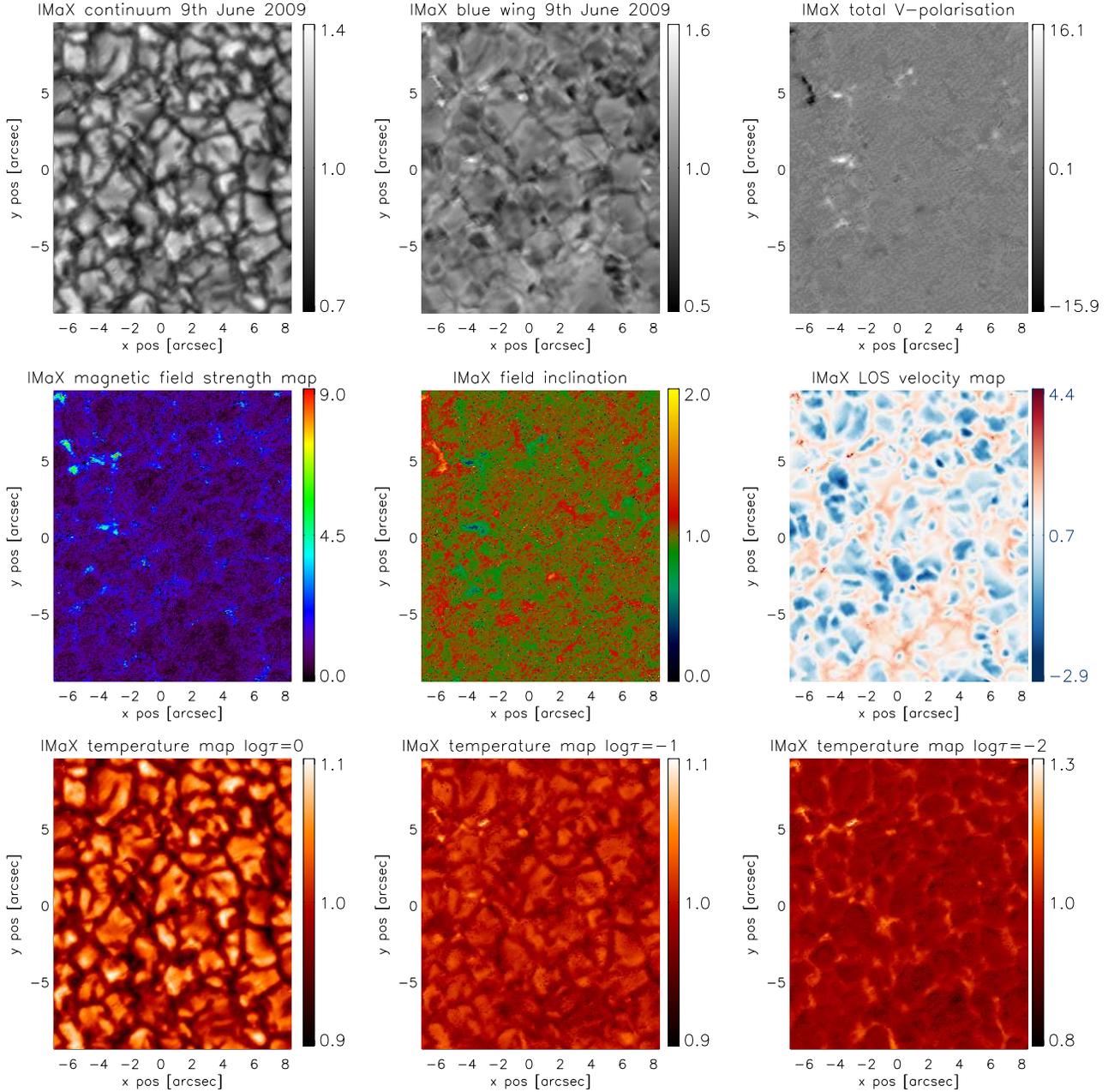}
\end{center}
\caption{Image examples (small detail of the full field of view) of the used data set and inversions. Shown in gray are the non-inverted data (first row, from left to right): a continuum intensity (227 m\AA) map; a blue-line wing ($-$40 m\AA) intensity map and a total abs(V) polarisation map (sign taken from the blue lobe of the iron line). Inversion results are shown in the second and third rows (from left to right and from top to bottom): magnetic field strength, inclination, line-of-sight velocity map (positive values correspond to downflows), and temperatures at log $\tau$=0, $-$1, $-$2. The color bars give the ratios of the quantities compared to the mean values computed over the complete field of view (except for the LOS velocity where the real values are stated in km/s). The mean values for the second and third row (inversion products) are: 0.15 kG and 90 deg for the second row and 6400 K, 5300 K, and 4900 K for the third row.\label{figure1}}
\end{figure*}

\section{Data}
The data were acquired with the IMaX instrument \citep[see][]{2011SoPh..268...57M} on board the \textit{Sunrise} balloon-borne mission \citep[][]{2010ApJ...723L.127S,2011SoPh..268....1B,2011SoPh..268..103B} which was flown from 2009 June 8 to 13 along a route from ESRANGE/Kiruna (balloon launching facility in northern Sweden) to Somerset island (northern Canada). During this period it ascended to the stratosphere to a height of about 37 km above the ground where it took practically seeing-free images of the Sun. The IMaX instrument is an imaging magnetograph making use of the Fe \scriptsize{\uppercase\expandafter{\romannumeral  1}} \normalsize 5250.2~\AA~line, which is the most magnetically sensitive line in the visible with an effective Land\'{e} factor of three. According to its instrumental paper \citep{2011SoPh..268...57M}: ``Gauss equivalent sensitivities are 4 G for longitudinal fields and 80 G for transverse fields per wavelength sample. The line-of-sight (LOS) velocities are estimated with statistical errors of the order of 5--40 ms$^{-1}$.'' This achieved performance of the instrument and used spectral line makes it the best bet to detect weak fields in quiet Sun when one does not have access to the near infrared part of the spectrum. In operation it takes at each of the $N_\lambda$ wavelengths, $N_p$ differently polarised (by using liquid crystal variable retarders as modulators) single exposures that are repeated $N_A$ times in order to increase the signal-to-noise ratio (S/N). Therefore, a single line scan data set consists of $N_\lambda\times N_p$ images that are taken in a time period equal to $N_\lambda\times N_p\times N_A\times t_{\mathrm{exp}}$, where $t_{\mathrm{exp}}$ is the exposure time of a single shot plus overhead for readout, etc. (about 250 ms).

The analysed data consist of 100 such single data sets obtained by the IMaX instrument in the so-called V5--6 mode. In this mode, V means that $N_p=4$ (vector mode), and 5 and 6 stand for $N_\lambda$ and $N_A$, respectively. The Fe \scriptsize{\uppercase\expandafter{\romannumeral  1}} \normalsize spectral line at 5250.2 \AA~was sampled at $-$80~m\AA, $-$40~m\AA, 40~m\AA, 80~m\AA~and 227~m\AA~off the line centre. The last sample is located in the nearby continuum. With this sampling scheme, the V5--6 mode reaches a typical S/N of 1000 and consists of all four Stokes parameters ($I$, $Q$, $U$, $V$).

The obtained sequence is broken in two parts: the first, lasting from 0:35:49 UT until 0:58:32 UT and consisting of 42 sets of images taken about every 33 s, and the second, with the same temporal resolution taken from 1:30:41 UT until 2:02:16 UT comprising 58 sets of Stokes profile maps. Both were acquired on 2009 June 9 close to the solar disc centre under exceptional quiet Sun conditions. The spatial sampling was 0.055 arcsec per pixel. Considering the detector area of $936\times936$ pixels$^2$, the field of view (FOV) is $51.48\times51.48$ arcsec$^2$. Due to the applied image reconstruction algorithms (resulting in apodised images; see \citealt{2011SoPh..268...57M} for details) the useful size of the FOV was reduced to about 780 $\times$ 780 pixels$^2$. The application of the phase diversity reconstruction algorithm yielded images with a typical signal to noise ratio of about 350 in the Stokes $I$ continuum and a spatial resolution of about 0.15 -- 0.18 arcsec.

A snapshot of a part of the field of view and some of the data products can be seen in Fig. \ref{figure1}. From left to right and from top to bottom, we display images of the continuum intensity, of the blue-wing intensity ($-$40 m\AA), of the total absolute circular polarization, of the magnetic field strength and inclination, of the LOS velocity, and of the temperature at log $\tau$ = 0, $-1$, and $-2$. The six latter images come from inversions with the SIR code \citep[Stokes Inversion based on Response functions;][]{1992ApJ...398..375R}.



\section{Analysis\label{analysis}}
We used an already reduced data set publicly available on the mission Web page. The full reduction comprises flat fielding, dark current reduction, polarimetric demodulation (necessary to retrieve the Stokes parameters from the measured linear combinations), cross-talk removal, and image reconstruction by using the point-spread function of the instrument as obtained from specifically devoted phase-diversity observations. All the detailed information about data reduction and reconstruction can be found in the instrument paper of \cite{2011SoPh..268...57M} and references therein. To stay as close as possible to the original data we have abstained from $p$-mode filtering. Hence, the $p$-modes may have an influence on the measured LOS velocities.

The inversions were carried out using two nodes for the temperature (i.e., modifying the initial guess atmosphere linearly) and one node for magnetic field strength, inclination, azimuth, LOS velocity, and microturbulence (assumed to be height independent). Due to the high spatial resolution, a zero macroturbulence velocity and a magnetic filling factor of unity were assumed. This means that for unresolved structures, the parameters returned by the inversion might represent averages over several atmospheric components. In total, seven free parameters were determined from the $5 \times 4 = 20$ observables. The average LOS velocity in the complete FOV was set to zero. Therefore LOS velocity values in this study have to be regarded as difference values to the average (quiet Sun) LOS velocity.

Several different algorithms for small-scale feature identification were developed in the past and presented in literature among with results on MBPs. To give at least two examples of automated and more sophisticated algorithms: a ``blob finding'' algorithm was successfully applied by \citet{1995ApJ...454..531B} on data of the Swedish Vacuum Solar Telescope for the identification of MBPs. In a more recent study, \citet{2010ApJ...722L.188C} devised an algorithm based on the idea of identifying MBPs within intergranular lanes. In a second step they use the fact that MBPs should be surrounded by pixels belonging to intergranular lanes and hence a ``compass'' search (investigation of the surrounding of a pixel) should yield only dark intergranular pixels. In addition other known properties of MBPs like their brightness and brightness gradient can be used. 

The MBP identification in this study was done by using the automated procedure developed by \citet{2009A&A...498..289U,2010A&A...511A..39U} on the blue-wing wavelength sample maps ($-$40 m\AA~from the line core). The high contrast of these images and the relatively high formation region of the line at that wavelength makes it ideal for MBP detection among the available IMaX data. The interesting features are situated in the intergranular lanes and are commonly associated with downflows, these downflows might further facilitate the detection by enhancing the contrast to the surroundings via line shifts (shifting the Fe line to longer wavelengths and thus giving larger measured intensities in the line wing). In contrast, the direct usage of the continuum intensity maps leads to identification difficulties, with a significant fraction of wrongly identified non-magnetic features \citep[][]{2013CEAB...37..459U}. We have refrained from using the high-contrast images provided by the Sunrise Filter Imager \citep[SuFI][]{2011SoPh..268...35G}, which are not affected by downflows and display MBPs clearly, because of its limited FOV, which would have restricted the statistical sample at our disposal too strongly. 


After the identification of the MBPs the tracking can be done \citep[for an overview and in depth discussion of several available algorithms; see][]{2007ApJ...666..576D}. The identification algorithm leaves us with a set of $x$ and $y$ positions of the intensity barycentre of the identified features (plus parameters such as brightness or size) for every image. The tracking procedure has now the purpose of connecting these single positional measurements of individual MBPs detected in each image to generate tracks or time-series of their evolutions. Such an algorithm must consider three constraints: a) the distance of connected features from one image to the next has to be smaller than a maximum distance given as $v_\mathrm{max}\times \Delta t$ where $v_\mathrm{max}$ is the maximum horizontal velocity of the MBP \citep[about 4 km/s; see, e.g.,][\citealt{2013A&A...549A.116J} who found MBPs to occasionally reach velocities of up to 16 km/s]{2003ApJ...587..458N} and $\Delta t$ is the temporal resolution of the data set (in our case 33 s); b) the set of connections between the feature positions should have a minimum total length, where the total length of a possible set of connections is defined as the sum of the distances between connected features; and c) no two paths of MBPs should cross each other from one time step to the next one as this would mean that one would have ``tunneled'' through the other one.



The approach to the tracking problem adopted in this paper does not describe the transition from one image to the next (executing a direct feature track assignment for every MBP), but instead, adds up additional pointer information. These pointers drive the programme to a follow-up and predecessor realisation. 
To obtain the pointers, we compare the current feature positions with the positions in the previous and subsequent image. The spatially closest following and preceding structure, regarding also the maximum distance rule, gets assigned to the MBP in question via two pointer variables which are added to the data structure. This structure contains then all the information of the single measurements, such as a unique identifier number (for the single measurement), brightness, size and so on, plus the pointer information to the follow up measurement and preceding one. After this tracking, the creation of time series of MBPs is a simple task. All features which have no pointer assigned to a previous realisation are obviously the starting points of MBP sequences. These structures can be taken and the evolution can be followed by jumping via the pointers from one measurement to the next. 


The retrieval of discussed physical quantities is now easily done by taking the plasma parameters at the barycentre position of the MBP. For a better comparison with the surrounding area and since the barycentre position did not always agree with the brightest or strongest pixel, we also computed the mean, maximum and minimum values in the vicinity of the feature's barycentre positions (keeping in mind that those evaluated positions must not coincide with the same pixel in all instants of time). The vicinity is defined as a square of 5 $\times$ 5 pixels$^2$ centered around the pixel of interest.

\section{Results\label{results}}
Prior to a statistical study of some 200 evolutionary tracks, we start this section with a detailed discussion of some specific, even though not classifying, cases (i.e., that each MBP track is too distinct from each other to be assigned to a specific class or type of MBP evolution). The first shown example is fairly representative of a canonical convective collapse. This theoretical concept is believed to represent an important process leading to the magnetic field amplification for small-scale magnetic structures, but it may not be the only possible amplification process. The second case illustrates that instead of the presence of a downflow per se, the interaction of the strong downflows with nearby small-scale upflows may be responsible for the magnetic field amplification, in agreement with recent findings by \citet{2014ApJ...789....6R}. In the third example we show an evolution in which plasma flows play only a minor role, if at all, for the evolution and creation of a strong magnetic field concentration. 

\subsection{Case I: double convective collapse event}
\begin{figure*}

\begin{center}
\includegraphics[width=0.82\textwidth]{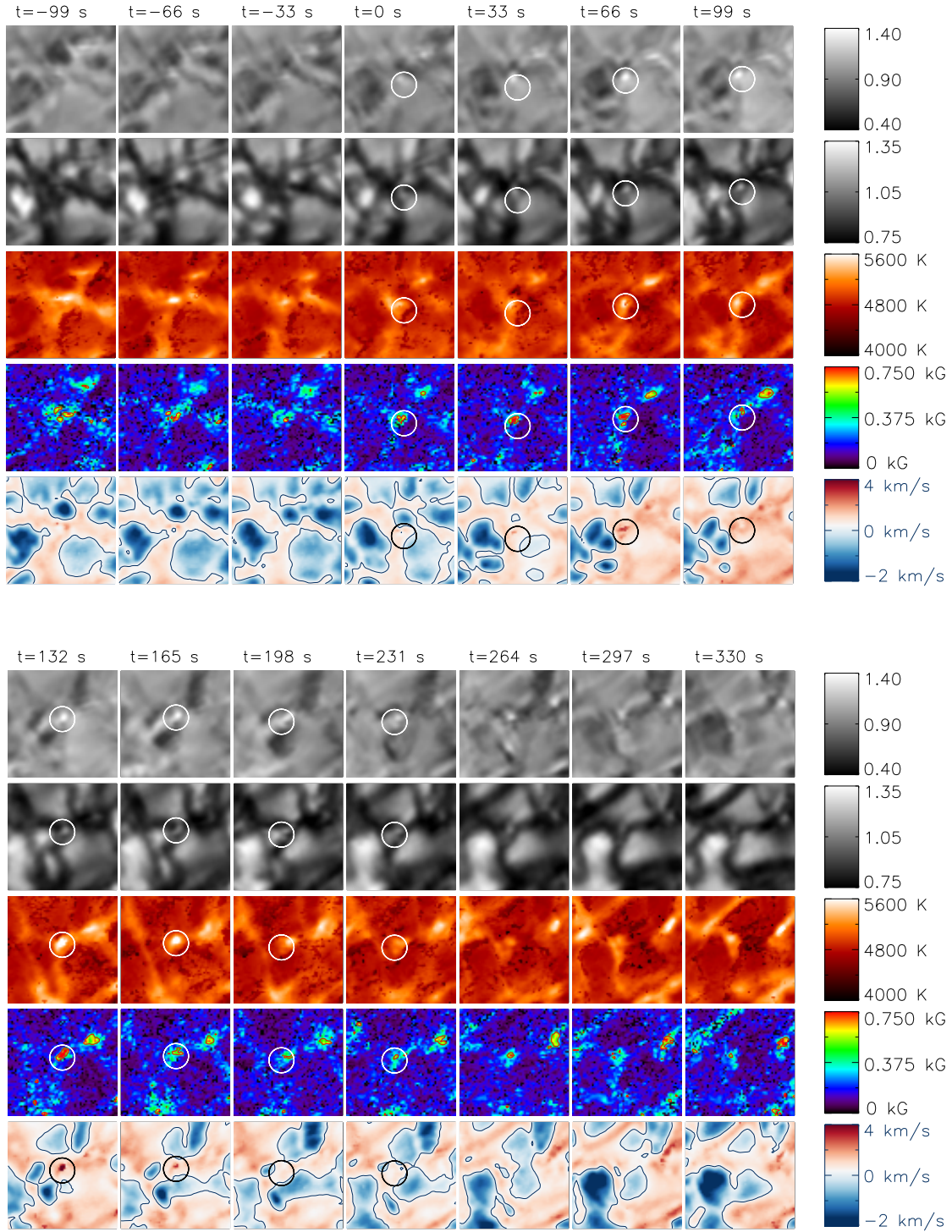}
\end{center}
\caption{From top to bottom: intensity image taken at $-$40 m\AA~from the Fe line core, continuum intensity image taken at +227 m\AA~from the line core, temperature map [K] obtained from the inversions at log $\tau=-2$, magnetic field strength map [kG], and velocity map [km/s]. The images are taken about 33 s apart from each other with the first three in the upper rows of panels and the last three in the lower rows showing the region of interest before and after the automated tracking. The tracked feature is marked by open circles. The FOV displayed is about 1.6 arcsec $\times$ 1.6 arcsec. The contours in the velocity maps correspond to the intersection between upflow and downflow region while the contour in the magnetic field strength maps outlines significant magnetic fields with strengths above 0.45 kG. The images have been clipped in such a way as to enhance the contrast with the clipping range given by the colourbars. \label{figure5}}
\end{figure*}

\begin{figure*}

\begin{center}
\includegraphics[width=1\textwidth]{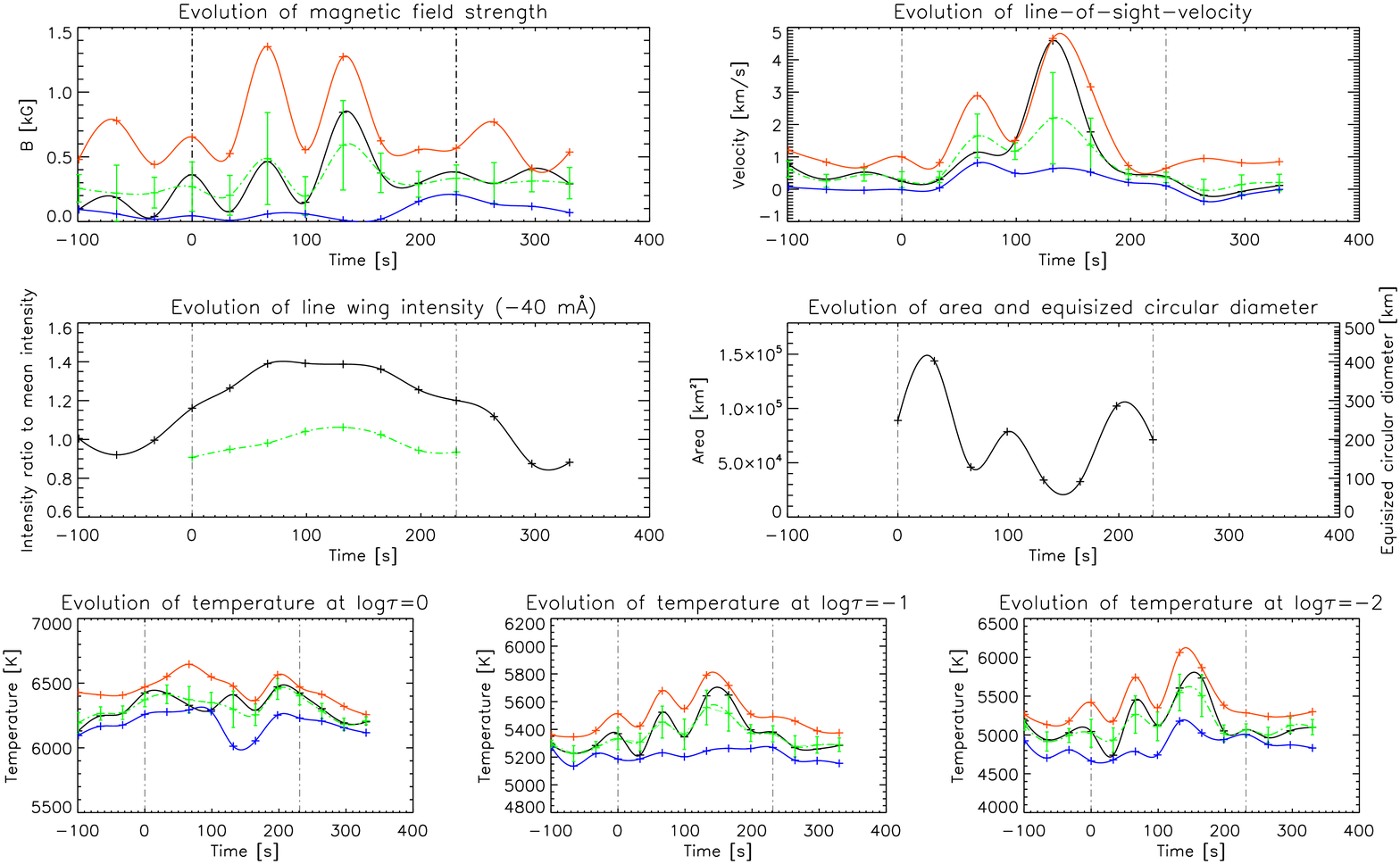}
\end{center}
\caption{First two rows: from left to right and top to bottom---evolution of the magnetic field strength, the LOS velocity component, the ratio of the line wing intensity at $-$40 m\AA~ at the barycentre to the average line wing intensity over the full FOV, and the feature size given in km$^2$ and as equisized circular diameter measured in kilometers. Bottom row from left to right: evolution of the temperature at log $\tau=0$, log $\tau=-1$, and log $\tau=-2$. The colour code is as follows: black lines show the quantities at the brightness barycentre of the tracked feature; orange/blue lines give the maximum/minimum values in a $5\times5$ pixels$^2$ box surrounding the barycentre; and green lines show the mean value in this subfield except for the line wing intensity plot. There the green line indicates the mean intensity over the segmented feature. Error bars correspond to rms values within the box. All lines are created by spline fits. The dash-dotted vertical lines indicate the starting and ending of the tracking period. The pre/post tracking evolution is shown for a period of about 1.5 minutes. \label{figure6}}
\end{figure*}

In Fig. \ref{figure5} we show the tracked feature in the plasma parameter maps. The first row gives the blue line-wing sample. The brightening is clearly seen in the maps and can still be visually identified in the first two images after the automated tracking has lost the MBP. This is probably due to a better human flexibility and ability in identifying structures and features compared to the employed automated algorithm. The next row illustrates the continuum sample. The MBP is formed at the merging site of several granules. The same feature can be identified also in the map of temperature at log $\tau=-2$ shown in the third row. The next row gives the magnetic field strength map. Here, the MBP is formed in a region of weak magnetic field that first strengthens then weakens again, and persists after the identification ceases. In the LOS velocity maps (fifth row) we see a strong downflow appearing at 66 s and again at 132 s. 
To probe the details of the MBP'´s evolution, we display in Fig. \ref{figure6} plots of the physical parameters within a square of 5 $\times$ 5 pixels centred on the barycentre of the structure in the blue wing maps. The evolution of the barycentre is plotted in black. Orange lines correspond to these pixels presenting maximum values of the corresponding quantity in the selected area. Blue lines correspond to the pixel having the minimum  value. Note that lines with the same colour may correspond to different pixels in each panel. Finally, dashed, green lines give the evolution of the average in the area. Error bars indicate the rms dispersion of each quantity within the selected box. The dashed vertical lines mark the beginning and the end of the automated MBP tracking.

Noteworthy, the brightness barycentre does not coincide spatially with the strongest field pixel, neither with the fastest downflow one. Except for a couple of instants where the latter is almost located at the brightness barycentre it seems clear that the maximum intensity and the largest velocities are located in the surroundings of the strongest field (tentatively, the magnetic structure'´s core). This is in agreement with the results found by \citet{2014ApJ...789....6R}, \citet{2014A&A...565A..84H}, and the ones outlined in the thesis of \citet[][]{Buehler} using a new inversion technique developed by \citet{2012A&A...548A...5V}. The authors of the latter analysed a plage region and came to a similar conclusion, namely that the downflows avoid the points of the strongest magnetic field.

Before the automated tracking starts it is clear that a magnetic field with a strength of a small fraction of a kG already exists. About 1 minute later, a first strong downflow of 3 $\mathrm{kms^{-1}}$ sets in that is accompanied by a clear strengthening of the magnetic field up to some 1.4 kG. Both, the downflow and the magnetic field strength, significantly decrease in just 30 s to 1~$\mathrm{kms^{-1}}$ and 0.5 kG just to rapidly increase again in the next half-minute. This time the field strength reaches 1.3 kG (0.9 kG for the brightness barycentre) and the downflow accelerates to almost 5~$\mathrm{kms^{-1}}$. Such rapid variations and fluctuations of plasma parameters were already reported previously by \citet{2013JPhCS.440a2032J}. While we see in these observations the magnetic field and LOS velocity maxima to occur co-temporal it was reported by \citet{2014A&A...565A..84H} that at the point of maximum field strength the LOS velocity cedes. The difference in observations and simulations might be explained by the lower temporal resolution in observations (about 30 s) compared to simulations (about 2 s).

        The size varies in a very nicely anti-correlated fashion with the field strength, as expected from the conservation of magnetic flux. At $\tau=1$, the temperature does not show clear correlations relative to the magnetic and dynamic evolution of the MBP (i.e., that the maximum field strength does not vary in phase with the temperature maximum), which is in agreement with the difficulties in MBP selection in the continuum maps. However, in higher layers (mostly at log $\tau=-2$), clear signatures of temperature enhancements can be seen at the same time as the convective collapse episodes happen. These observational findings agree mostly with recent simulations \citep[see Fig. 1, fourth panel, of][]{2014A&A...565A..84H} where an increase in temperature in all three discussed atmospheric layers have been seen.

Still within the tracking period and only 33 s after the second collapse, both, the magnetic field strength and the plasma flow, relax to the values they had prior to the first convective collapse. Therefore, the kG state does not seem to be a stable configuration of this feature that has nevertheless experienced a double convective collapse event.

\subsection{Case II: illustration of the importance of shear flows/upflows}
\begin{figure*}
\begin{center}
\includegraphics[width=0.82\textwidth]{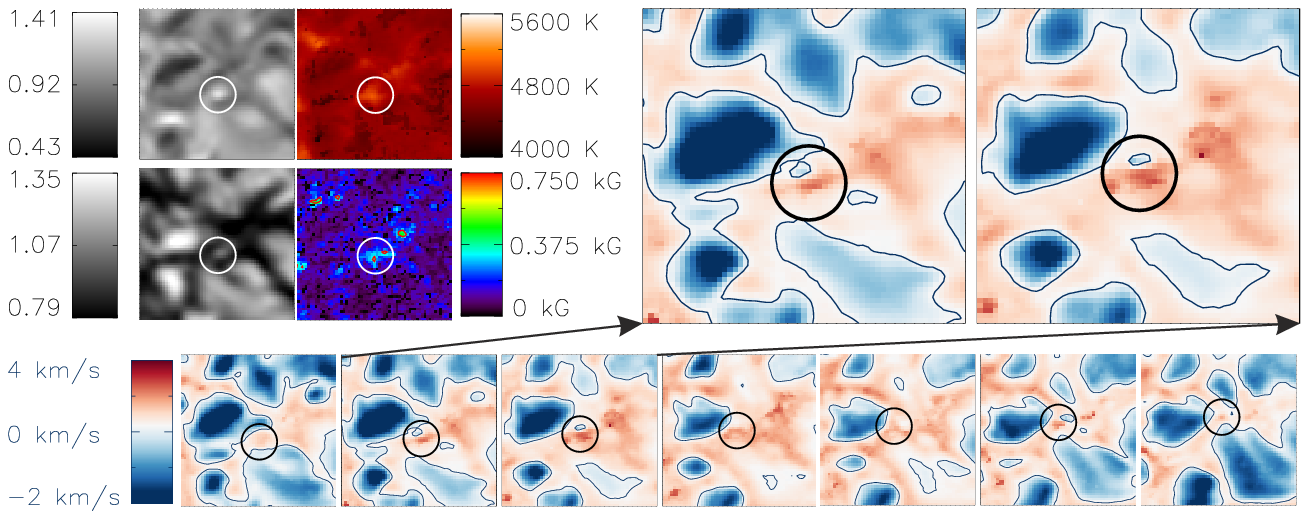}
\end{center}
\caption{Similar to Fig. \ref{figure5} but shortened to the crucial aspects during this particular MBP evolution where isolated upflows appear in the downflow lane next to the magnetic feature. The upper left corner panel shows in four subplots the initial conditions of the surrounding of the tracked MBP (from left to right; top to bottom): blue line-wing intensity, temperature response in the log $\tau=-2$ layer obtained by inversions, continuum, and magnetic field strength map. Below: the complete tracked evolution of the feature in the LOS velocity maps is shown. Right top corner enlargement of two time instances showing a co-temporal and co-spatial small-scale upflow. The LOS maps are taken 33 seconds apart from each other.\label{figure7}}
\end{figure*}

The maps of the plasma parameters and the evolution of a second MBP are depicted in Fig. \ref{figure7}. While the track of the feature in the blue line-wing, the continuum, the temperature, and also in the magnetic field strength map does not show any special behaviour with respect to the other studied cases, one can see an interesting detail in the velocity maps. Starting in the second image after the automated tracking started, and vanishing again in the fourth image, there is a very small intergranular upflow in the middle of the downflowing region, very close to the detected barycentre of the MBP. The connection of the upflow with the neighbouring granule and its coincidence with strong downflows, have recently been investigated by \citet{2014ApJ...789....6R}. In addition such small-scale upflows can be seen in simulations like the one of \citet[][e.g., Fig. 4, middle panel and lower panel]{2014A&A...565A..84H} where one can see that the barycentre of the found MBP is situated actually just in between a strong downflow and an upflow.

Interestingly, within the same evolutionary track, we find a second such occurrence in the last two tracked images. Again, a small upflow evolves very close to the studied downflow and disappears just after one more image.

\begin{figure*}
\begin{center}
\includegraphics[width=1\textwidth]{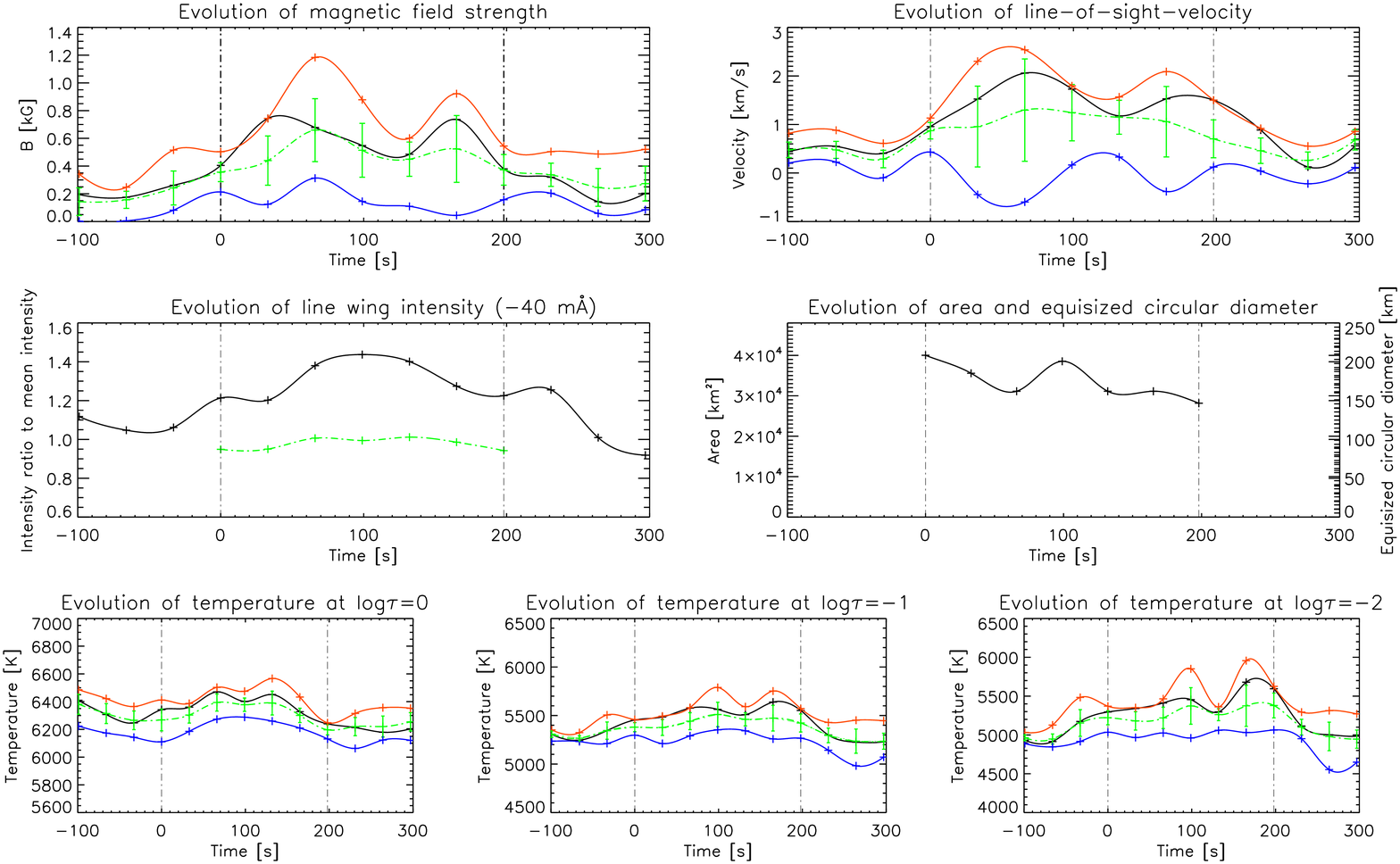}
\end{center}
\caption{Same as Fig. \ref{figure6} but for the feature shown in Fig. \ref{figure7}.\label{figure8}}
\end{figure*}

As for the former case, Fig. \ref{figure8} illustrates the evolution of the brightness barycentre, and of the maximum and minimum quantities as well as the average over 5 by 5 pixels$^2$ centred on the barycentre. Again, this MBP shows two conspicuous magnetic strengthenings. The strongest pixel reaches 1.2 kG and later 1 kG, while the barycentre itself has some 0.8 kG on both occasions. Peaks in the downflow of 2.5 and 2 kms$^{-1}$, respectively, appear simultaneously with the magnetic field strength maxima. Noteworthy is the fact that at the time of the strong downflows, two upflows are detected within the defined vicinity box around the MBP barycentre. These upflows cover only a rather small region of approximately 10.000 km$^2$ (estimated by visual inspection) and are weaker than the downflows, but strong enough to be reliable with values between $-0.5$ to $-1$ kms$^{-1}$. These striking coincidences of up- and downflows, while the magnetic field strengthens, point to the idea that both processes might be coupled somehow and that the canonical convective collapse onset could be accompanied by some mechanisms \citep[e.g.,][]{1998ApJ...495..468S,1999ASPC..184...38S,2011ApJ...730L..24K,2014ApJ...789....6R}.
The other plasma parameters show no behaviour worth mentioning except for the evolution of the temperature in the higher photospheric layers. Interestingly, this temperature displays a double peak, as seen for the magnetic field strength and velocity evolution, but only the second peak is co-temporal to the velocity and magnetic field strength peaks.

A manual inspection of all the tracked MBPs show that about 10\% of all the tracked features show stronger spatially located upflows close to the downflows related to the magnetic feature. Therefore the discussed case might still be quite a special one but definitely not the only one in which a spatially correlated upflow plays an important role in the evolution of the feature.

\subsection{Case III: magnetic field evolution without strong LOS velocities}
\begin{figure*}
\begin{center}
\includegraphics[width=0.82\textwidth]{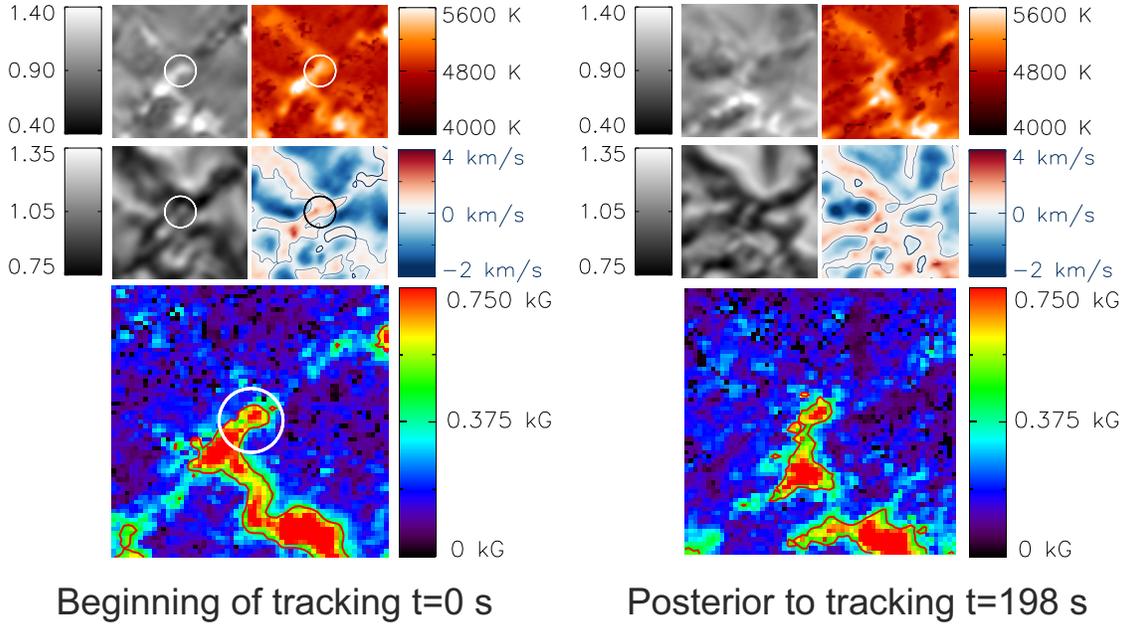}
\end{center}
\caption{Shortened version of Figs. \ref{figure5} and \ref{figure7} for a feature track which shows no strong influence of up/downflows on the strength of the magnetic field. Only the first tracking instance and the first one posterior to the tracking are shown. \label{figure9}}
\end{figure*}

The third presented case complements the picture and shows the complexity and variety of possible evolutionary paths. This case differs from the others by the large extent of the magnetic field patch at whose border the MBP feature was detected, as can be seen from Fig. \ref{figure9}. This patch also harbours a number of further bright points and filigree and is probably a network magnetic field concentration which is a major difference to the other, more or less well isolated, small and compact, MBPs.

\begin{figure*}
\begin{center}
\includegraphics[width=1\textwidth]{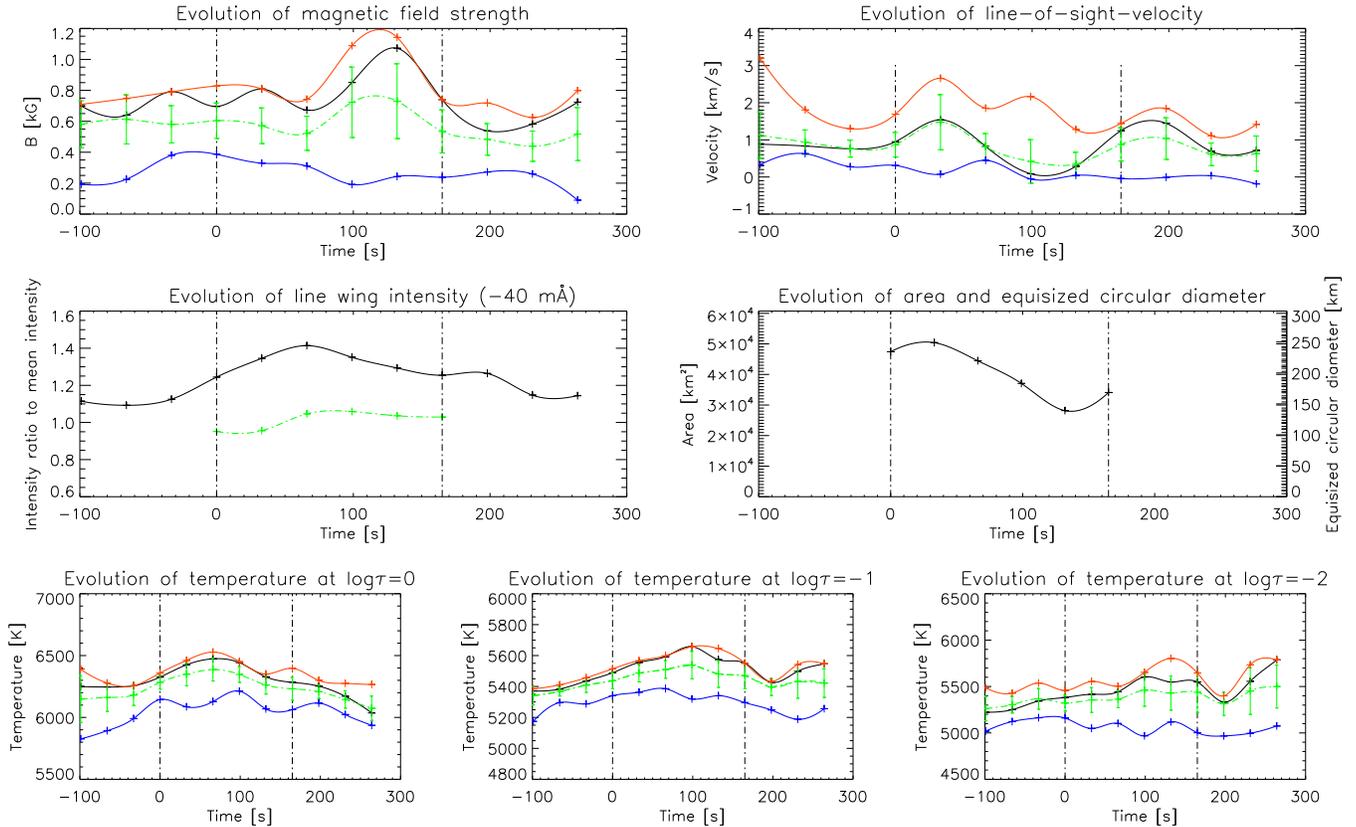}
\end{center}
\caption{Same as Figs. \ref{figure6} and \ref{figure8} for a feature track which purports to show no strong influence of up/downflows on the strength of the magnetic field. 
\label{figure10}}
\end{figure*}

Figure \ref{figure10} illustrates the evolution of the different plasma parameters. The magnetic field strength shows a well defined maximum, which exceeds the kG threshold, 132 s after the start of tracking. Also, the size follows the commonly accepted picture of a fast shrinkage with a co-temporal increase of magnetic field strength \citep[see e.g.][]{1979SoPh...61..363S,2009A&A...504..583F}. But the most interesting curve for this case is the evolution of the LOS velocity component. There is a strong downflow of about 2.6 kms$^{-1}$ at the beginning of the tracking witnessed by the orange line (maximum in the FOV) that gets weaker (dropping to below 0.3 kms$^{-1}$ for the barycentre - black line - and average values - green line) when the magnetic field strength maximum is reached. It increases again with weaker magnetic fields when the tracking is, in fact, stopped. This is in contrast to the other studied cases, where the LOS velocity component behaved more or less co-temporal with the changing magnetic field strength. On the other hand this is in perfect agreement with the simulations done by \citet{2014A&A...565A..84H} who stated that the LOS velocity cedes or at least strongly decreases at the moment of intensity and magnetic field maximum. In their simulations the LOS velocity peaked always before the intensity and magnetic field strength reached its maximum. An increase in the temperature of the higher layers can also be seen, although weaker than in the two former cases.

The major difference to the two former cases is probably that this feature was not created out of a weak internetwork magnetic field patch but is on the contrary related to an extended network magnetic field region and second that the formed bright point was never really isolated. Even though it has been partly dislocated and split up from the network patch. Furthermore the final field strength at the point of dissociation is much higher with roughly 0.6 kG. This makes it more compatible to the two cases discussed in the Appendix of the paper of \citet{2014A&A...565A..84H}, where the MBPs were formed in a magnetic field patch already clearly exceeding at the point of formation the 1 kG threshold. Besides, the tracking in such ``network'' case is more elaborate as nearby features constantly interact with the feature under investigation. We will go into more details on this in the discussion section.

From the three presented examples, it is clear that while one can learn a lot from single (partly special) cases, it is quite hard to understand the evolution of MBPs on average or in a general way. Different cases seem to be just too distinct from each other to generalise based on such singular observations. To achieve the goal of a more general characterisation we applied statistical methods on all of the tracked and investigated evolutions at hand.

\subsection{MBP evolutionary statistics}
\begin{figure}
\begin{center}
\includegraphics[width=0.45\textwidth]{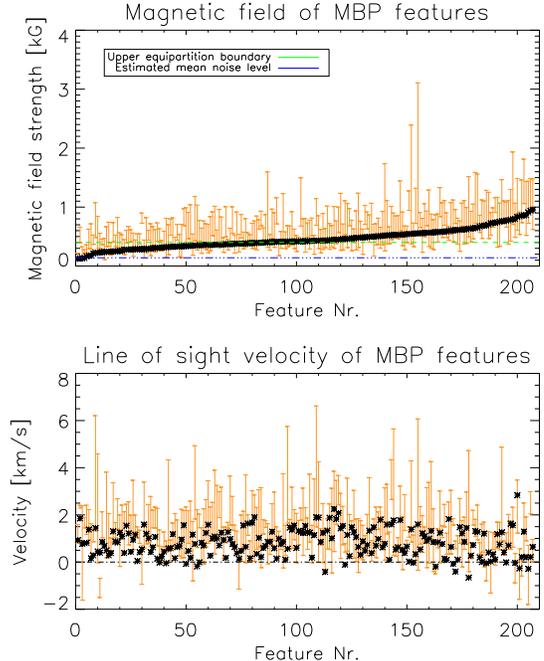}
\end{center}
\caption{Top panel: the initial magnetic field strength of all the tracked MBPs sorted according to an increasing initial field strength from left to right (black asterisks). The vertical bars of the plot give the final field strength (just after tracking ends; lower end of bar) and maximum field strength reached during their temporal evolution (upper end of vertical bar). The blue dash-dotted line illustrates the mean noise level (0.14 kG; estimated via averaging over pixels with polarisation signals $<$~3~$\sigma$) in the field of view. The upper boundary of the equipartition field strength range is given by a green line (0.4 kG).
 Bottom panel: the LOS velocity, with positive values representing downflows, for the same features as in the upper panel, plotted in the same order. The asterisks show the initial LOS velocity while the vertical bars give again the final and maximum downflow value of the velocity obtained during the lifetime of the features. All the values were taken at the barycentre position of the MBP.
\label{figure11}}
\end{figure}

In this subsection we outline the major statistical results derived from the roughly 200 single MBP tracks, incorporating a few thousand single MBP features identified in the course of this study. For this purpose, we have sorted the tracks with increasing initial magnetic field strength, whereby the field strength prior to the first tracked position is used. In the top panel of Fig. \ref{figure11} we plot this field strength (black asterisks) together with the maximum field strength reached during the evolution of the tracked MBPs (top of the vertical bars) and their dissolution magnetic field strength (bottom of the vertical bars; that is extracted at the last known position in the first image after the tracking).

MBPs are most likely distinctive features that should have a different magnetic field strength distribution---and, hence, also onset and dissolution strengths---than the pure background magnetic field in the FOV due to noise in the polarisation signals. To estimate such a background noise field strength we have averaged that of pixels having Stokes $Q$, $U$, and $V$ signals below $3\,\sigma$ of the continuum intensity and found $0.14 \pm 0.08$ kG. This noise level is displayed in the top panel of Fig. 8 with a blue, dash-dotted line. Since 97 \% of the tracks show at least at one instance of time field strengths larger than 0.38 kG, that is, the average background noise field strength plus three times the rms, we make sure that the overwhelming majority of the automatic MBP detections correspond to distinctive magnetic features. 
The previously mentioned threshold strength of 0.38 kG is almost coincident with the upper limit (0.4 kG) for the equipartition field strength range \citep[see, e.g.,][and references therein]{2010ApJ...723L.185W} which is marked with a green, dashed line in the upper panel of the figure. A glance at the figure reveals that the onset field strengths of all the MBPs are distributed around that equipartition value. Indeed, the median of these onsets is 0.42 kG.
The vertical bars show us that most of the MBP field strengths stay below a deduced field strength of one kG with an amplification of two to three times their onset value. Most of them dissolve with field strengths below the equipartition value. 

The LOS velocity component (second panel of Fig. \ref{figure11}) displays in general downflows, for both, the onset and the dissolution, i.e., there are only a few cases of detected flow reversals at the very end (24 upflows out of more than 200 features resulting in about 10\% of measured flow reversals). Important to note is that the statistics is done for the barycentre position and not for the maxima or minima in the vicinity of the MBP (those quantities have been shown previously in the evolutionary track Figs. \ref{figure6}, \ref{figure8}, and \ref{figure10} in addition to the barycentre), i.e., it is still possible that there are even more upflows in the vicinity. Also, because MBPs are located almost exclusively in downflowing intergranular lanes, it may well be that upflows are hidden by the effects of scattered light. Illustrations on how scattered light can hide small-scale flows of opposite sign to the dominant ones are given by \citep{2011ApJ...734L..18J,2011Sci...333..316S} for sunspot penumbrae. The initial phase is characterised by weak downflows on the order of 1 km/s ($0.8\pm0.6$ km/s), which is also the case for the dissolution stage when plasma flows are nearly back to the initial ones (only slightly changed to an even smaller average value of $0.6\pm0.6$ km/s). In a few cases it can be observed that the LOS velocity component changes sign, meaning that there is an upflow at the very end of the evolution. In about half of the cases, the downflows only reach a maximum value of about 2 km/s and only a few cases (about 5 \%) show very strong downflows exceeding 4 km/s, with the strongest reaching 6--7 km/s.


\section{Discussion\label{discussion}}
In the following we discuss the various aspects of this work, evaluate it critically and compare the methods and results with the literature. We structure the discussion by breaking it into subsections that consider the various steps of the analysis and the results. 
\subsection{Identification of MBPs in the blue line-wing:}
Narrowband filtergrams taken in the G band, which is a CH absorption band centred around 430.5 nm, have become a standard observable in recent MBP studies \citep[see, e.g.,][]{2004A&A...428..613B,2010ApJ...715L..26S}. Since the formation height of the G band is at mid-upper photospheric layers, the closest available proxy within the data set at hand is one of the wavelength samples closer to the line core which is therefore more sensitive to the temperature in such layers. The very high temperature dependence of the Fe \scriptsize{\uppercase\expandafter{\romannumeral  1}} \normalsize 525.0 nm line strength makes this a sensitive parameter. Hence, we have chosen the blue line-wing sample at $-40$ m\AA ~off the line centre for the identification of MBPs. While this facilitates the detection of MBPs as they show most of the time downflows which contributes to an even higher intensity into this line sample (due to line shifts), it bears the risk of masking upflows as those velocities might decrease the brightness in the blue line-wing sample.

\subsection{Feature tracking:}

\begin{figure}
\begin{center}
\includegraphics[width=0.4\textwidth]{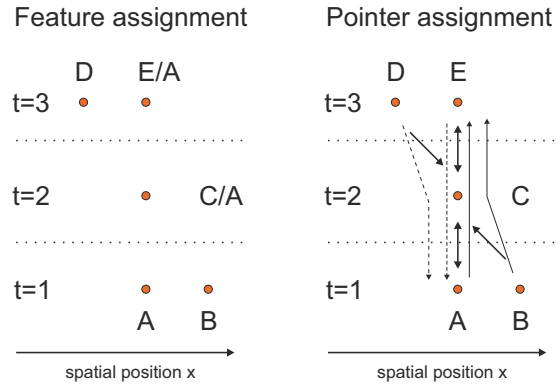}
\end{center}
\caption{Illustrating a tracking example. On the left side the feature assignment approach is shown while the right side depicts the pointer approach. The solid line arrows depict the pointer connecting instances of MBP realisations with each other while the thin solid and dashed line shows complete found tracks forward in time and under reversed time conditions. The advantages and disadvantages of both approaches are discussed in the text.
\label{figure12}}
\end{figure}

Two contrasting tracking approaches are illustrated in Fig. \ref{figure12} on the basis of a simple example covering three time steps, with a maximum of two features per time step. On the left-hand side we have a feature assigning tracking approach. Imagine that we have two features (A and B) at a certain instant. If at the next one we only find one feature (C), which is closer to the old position of A than to the old position of B, then we will replace the MBP identifier C by A (always under the prerequisite that the features are close enough, determined by a maximum distance variable). The same happens in the following time step with feature E, which is most likely not a new structure but the continuation of A. On the right-hand side of Fig. \ref{figure12} we have the pointer solution to the tracking problem. Here we are not replacing the feature identifier with a general track identifier and therefore assigning the measured MBP to a certain track. Rather, we add two pointers to every feature pointing to its predecessor and its follower. In this case one measurement can be used by several tracks (several other MBPs point to it) but overall the algorithm remains fairly simple by only allowing a given MBP to point to another one, namely, the spatially closest. For a more general discussion and the implications raised by other tracking approaches we refer to the publication of \citet{2007ApJ...666..576D}.

The feature assigning tracking has the advantage of creating a time reversible solution but the assignment of the feature track can be very complicated if there are several features close by in one or more instances giving rise to plentiful possible combinations and hence potential feature tracks. This is particularly true if the only criterion the decision is based on are the locations of the barycentres as in this work. More sophisticated tracking algorithms (but slower in runtime) would also incorporate the shape of the tracked features and look for example on the overlap between two realised instances of the feature \citep[e.g.,][]{1999ApJ...511..932H}. On the other hand the pointer tracking approach is fast and easy to realize but not reversible as the example, depicted in Fig. \ref{figure12}, shows. The thin lines on the right-hand side subfigure show the realised MBP tracks. Clearly one of the solid line tracks (B-C-E) is different to the dashed line track (D-C-A). This means, if the time would be reversed, different features would get connected to each other. 
At the end, especially for our case, it is not such a serious problem as we are i) in quiet Sun and hence can expect that there are mostly isolated MBPs and ii) we are not going to estimate parameters which could be crucially influenced by this behaviour. Such parameters would be number densities, fractional occupation rates of the surface (here double counting of features would lead directly to errors), flux emergence or decay rates 
and of course lifetimes. 


\subsection{Inversions:}
The used data are among the best currently available spectropolarimetric data sets with regard to their temporal and spatial resolution and in particular the consistently high data quality over the whole time series. Furthermore, the spectral resolution is sufficient to provide enough information for standard inversions such as Milne--Eddington (M-E) inversions. In our case the inversions were carried out with the SIR code \citep[see][]{1992ApJ...398..375R} under the assumption of an easy atmospheric model stratification, namely assuming two nodes for the temperature (that is, linear perturbations), and one node (constant perturbations) for the three components of the magnetic field, the LOS velocity, and the microturbulence.
Therefore, only seven free parameters are used, even less than with M-E inversions, but gaining additional information on the temperature stratification. Due to weak Stokes $Q$ and $U$ signals and the general tendency of inversion codes to overestimate magnetic field inclination in the case of such weak Stokes signals \citep[see, e.g.,][]{2003A&A...408.1115K,2011A&A...527A..29B,2012A&A...547A..89B} we refrained in this work from analysing and discussing the evolution of magnetic field inclinations.

A closer look on the results of the inversion shows that the formal error of the inversions for the inclinations is in most cases about 25 degrees but in some very poorly defined cases (practically there is only noise in the $Q$ and $U$ component) this error can be well above 40 degrees. Some pixels on the other hand, with higher signal strengths, show formal inclination errors of less than 8 degrees.

The reliability of the SIR inversion results on these data for other plasma parameters was already assessed and reported by \citet{2014ApJ...789....6R} who estimated typical uncertainties (rms) of 0.1~kG and 100~ms$^{-1}$ from several different realizations of added noise to the data at a level of 10$^{-3}$ ${I_c}$. In detail they repeated the inversions with 100 different realisations of noise and were thus able to assess the impact of noise on the quality of the inversions with the previously stated outcome. Hence we take these estimates also as valid for our analysis as we work with the same data set as well as with the same inversion code set up in the same fashion. Nevertheless, one could legitimately wonder whether there are limits above which the inferences are wrong, based on the typical behavior expected for Stokes $V$ in the two extreme cases of weak and strong fields. These  doubts could be especially addressed to the strongest fields in our observations that might be too strong. The errors might come, e.g., from the quite large instrumental broadening on a fairly reduced sampling. We are, however, confident on the Requerey et al.'s estimates. 

To digress a little on this, let us point out that, first, SIR already takes the broadening into account and, second, that arguments based just on Stokes $V$ may fail because the inversion code makes full use of the information contained in all four Stokes parameters. Information from Stokes $I$ is also important and even the fact that Stokes $Q$ and $U$ are typically small can be relevant to the final fit. Hence, our only intuition that is built on Stokes $V$ may not be of good help. Last but not least, the most reassuring confirmation is already present in our statistical results. Note that field strengths in Fig. \ref{figure11} barely exceed 1.5 kG. Among the approximately 2000 analyzed sets of profiles, only two of them display clearly erroneous values above 2 kG. We have neither eliminated nor analyzed in detail these two singular cases. On the one hand, showing them on the plot stresses the good quality of the results. On the other hand, a manual repair after a specific analysis is indeed irrelevant for our statistical purposes.


\subsection{Comparison with previous studies:}
\cite{2001ApJ...560.1010B} were among the first ones to report on the observational detection of a convective collapse \citep[earlier observations were done already by, e.g.,][]{1999ASPC..183...36S,1999ApJ...514..448L}. 
They studied exhaustively the observed Stokes profiles and performed also an inversion of the spectropolarimetric data. The occurrence of a convective collapse with magnetic field strengths increasing from 0.4 kG to 0.6 kG and downflow velocities of up to 6 km/s were confirmed by them. The overall process lasted for about 4 minutes before a seemingly rebounding plasma created a strong flow and shock which finally seemed to disintegrate the structure. One might ask at this point if it really takes a full-fledged instability to produce this moderate increase in magnetic field strength or if other mechanisms could also be responsible for moderate increases in the magnetic field?

\citet{2008ApJ...680.1467S} and coworkers were among the first to use the recent \textit{Hinode} spacecraft and the onboard spectropolarimeter to investigate high-speed local mass downflows. They prepared a detailed study of the Stokes profiles taken in several interesting regions from the sunspot moat to the quiet Sun. In the case of the quiet Sun they were able to identify co-temporal brightenings in the G-band and in the Ca \scriptsize{\uppercase\expandafter{\romannumeral  2}-H}\normalsize filtergram occurring together with strong transient downflows. \citet{2008ApJ...677L.145N} was focusing then on such downflows in the quiet Sun and reported a detailed case study of a convective collapse and the formation of an MBP from the \textit{Hinode} spectropolarimeter data. The inversion of the data yielded the magnetic field strength and the LOS velocity component. Furthermore, the brightness evolution was followed and, from the brightness difference, an estimate of the temperature difference was obtained. As in the study of \cite{2001ApJ...560.1010B} they found an increasing plasma downflow which went along with increasing magnetic field strength and brightness. The maximum field strength was reached about 3 minutes after the onset of the downflow with a value of 2 kG representing an amplification factor of four. The maximum downflow velocity was reached at the same time, with a value of 6 km/s. Afterwards, the velocity decreased while the brightness still increased for about 100 more seconds, with the magnetic field strength remaining nearly constant. The final point of the observed evolution was reached when the velocity reversed and an upflow of about 2 km/s set in, causing the magnetic field to reduce to about 1.5 kG along with a reduction of intensity. Unfortunately, the authors did not follow the evolution any further. We note that in this exceptionally well-studied evolution of an MBP the maximum field strength reached is near the upper limit of observed field strengths for MBPs. It did not take the community long to reproduce the results of \citet{2008ApJ...677L.145N} on a broader statistical base which was done by \cite{2009A&A...504..583F}, who studied a total of 49 such convective collapse events with spectro-polarimetric data and filtergrams of Mg \scriptsize{\uppercase\expandafter{\romannumeral  1}-b$_2$}\normalsize as well as Ca \scriptsize{\uppercase\expandafter{\romannumeral  2}-H}\normalsize (higher photosphere and lower chromosphere). They were able to confirm the major implications of the convective collapse scenario; A shrinkage in size with observed downflows and the forming of bright points in the photosphere as well as in the chromosphere for most of the observed features.

The previous studies of \citet{2001ApJ...560.1010B} and \citet{2008ApJ...677L.145N} are congruent in reporting strong downflows (of about 6 km/s), followed later on by an upflow, which disintegrated in the first case the magnetic field, while it weakened the field in the second. Indeed we were able to identify such strong downflows in several cases, e.g. in 14 tracks the downflow exceeds 4 km/s (about 7\% of the total amount of studied evolutions) and in five cases (about 3\% of the total number of tracks) the value of 6 km/s is exceeded (or at least closely reached). Nevertheless this did not lead to an observable velocity reversal with upflows and a weakening of the magnetic field strength of MBPs. 
All the same, some upflows have been observed very close to our MBPs at given moments of their evolution. Our identification algorithm---with different criteria to that of \citet{2008ApJ...677L.145N}---or different amounts of magnetic flux contained in the structures could possibly explain the discrepancy. The identification in the blue line-wing could be ruled out as an explanation for the discrepancy as upflows would increase the probability of ending the tracking of a feature (causing lower brightnesses in the blue line-wing) and hence be in favour for detecting upflows at the end of the tracking period. Another possibility would be that the better spatial resolution reached in our study reveals spatially separated up and downflows where previous studies only found a mixture of both, resulting in an effective up or downflow depending on what quantity had a stronger influence on the average.

In a recent study, using the Imaging Spectro Polarimeter \citep[CRISP; see][]{2008ApJ...689L..69S} at the Swedish Solar Telescope, \citet{2011A&A...529A..79N} presented results about strong transient downflows. These downflows were obviously associated with the creation of strong magnetic fields (flux tubes) and hence interpreted as manifestations of convective collapse. The author reports eight cases with downflows peaking at a maximum between 3.0 km/s and 5.2 km/s and showing peak magnetic field strengths from 0.48 kG to 1.32 kG. As in our study, the author does not find a reversal of the downflow into an upflow to cause the dispersion and weakening/disintegration of the flux concentration as reported by others. Furthermore the magnetic concentrations found by Narayan are of a transient nature with short lifetimes and do not lead to a permanent increase in magnetic field strength, in contrast to the MBP followed by \cite{2008ApJ...677L.145N}. Overall our findings in the current study agree very well with the ones presented in the paper of Narayan. Furthermore, we are able to identify small concentrated strong upflows close to downflows in Narayan's study (as reported here for our second case) for his case g (Fig. A.6, first row fifth and sixth image within the marking circle), even though it looks in the particular case as if this would be the starting point of a newly born granular cell, and for his case h (Fig. A.7, first row, second to fourth image left of the marked region).


In another recent study, \citet{2014ApJ...789....6R} investigated in detail the evolution of a single magnetic structure from its formation due to flux concentration by granular advection, the further amplification by a process interpreted as convective collapse,
till an oscillatory-like behavior at the final state of a collapsed flux tube is reached. During the convective collapse process they were able to observe the formation of two MBPs which were located close to the edge of the original magnetic structure which undergoes the collapse. Furthermore they showed that the plasma parameters of these features are correlated in a complex and not straightforward manner to each other (undergoing rapid fluctuations without a clear correlation to the brightness of the MBP) while the core of the magnetic structure (which undergoes the collapse) shows a nice correlation between magnetic field strength and LOS velocity. These findings may explain the complexity seen in the evolutionary tracks of the plasma parameters (Figs. \ref{figure6}, \ref{figure8}, and \ref{figure10}).

Finally, our study agrees with recent computer simulations such as the one of \cite{2010A&A...509A..76D} where the authors compare the magnetic field amplification process as seen in MHD simulations with those observed by \textit{Hinode}/SOT data. They found downflow velocities between 5 km/s and 10 km/s in the simulations. For the strongest of the studied downflows they observed a velocity reversal after 200 s. However, the upflow did not disintegrate the magnetic field concentration, in contrast to the findings of \cite{2001ApJ...560.1010B}, but led to a considerable weakening of the field as was observed by \cite{2008ApJ...677L.145N}. An interesting recent study shedding light on the evolution of MBPs is the one of \cite{2014A&A...565A..84H}, which shows nicely the creation of strong kG magnetic field brightenings in numerical simulations. The major difference to most observational papers is the non-co-temporality between the magnetic field strength, brightness, and LOS velocity maximum. Only our third observational case agrees with their general finding of preceding velocity maxima before field strength and intensity maxima are reached. Besides, they see moderate downflows with 3 to 4 km/s leading to really strong magnetic fields exceeding even the 2 kG threshold. In one of their three cases they are able to identify a very small flow reversal but generally speaking the evolution of the MBPs is dominated by downflows. Interestingly the brightness maxima of their first case lies between a downflow and a weaker upflow like in the second presented case of the study at hand.

If we compare the properties of our features, which are mainly located in the internetwork quiet Sun, with the results of high-resolution studies of magnetic elements in active region plage, we see clear differences. For example, our magnetic features display often sizable downflows over most of their generally short lifetime (at least as defined as their visibility as bright points), see Figs. 8, 11, and 12. The magnetic features in plage regions are surrounded by strong downflows, but in their interiors they tend to be relatively free of downflows, as found by \citet{2005A&A...435..327R}, \citet{2007ApJ...655..615L} and \citet{Buehler}. Obviously, these plage magnetic elements have a rather different internal structure than the highly dynamic internetwork structures that we are studying.

\begin{figure}
\begin{center}
\includegraphics[width=0.45\textwidth]{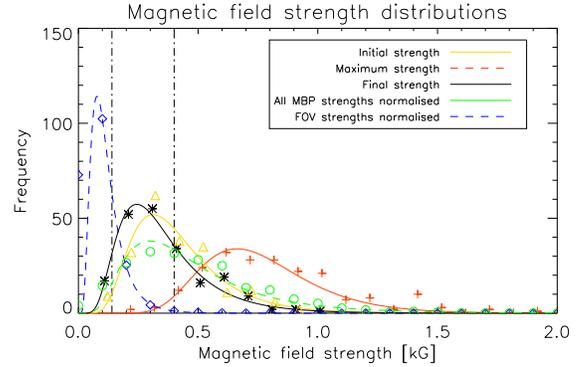}
\end{center}
\caption{Magnetic field strength distributions of the maximum MBP field strengths (orange symbols), the initial magnetic field strengths (yellow), and the dissolution field strength (black). The distribution for the complete FOV and for all MBP measurements, both normalised to the number of tracks, are shown (blue and green symbols, respectively). All MBP values were taken at the barycentre position of the identified MBPs at various time steps. The left dash-dotted, vertical line indicates the magnetic field strength (0.14 kG), corresponding to the average background noise and the right one marks an upper limit for the equipartition field strength (0.4 kG). Solid and dashed lines represent log-normal fits through the measured distributions.
\label{figure13}}
\end{figure}

\begin{figure}
\begin{center}
\includegraphics[width=0.45\textwidth]{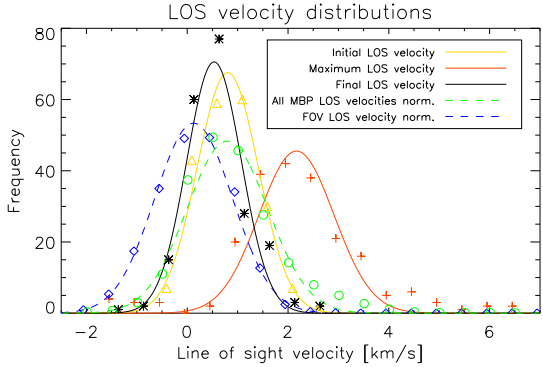}
\end{center}
\caption{Line-of-sight velocity distributions of MBPs in different evolutionary stages as well as the distribution for the whole FOV. Initial velocities are in yellow, maximum velocities are in red, and final values during detection are in black. The used values were taken from the barycentre position of the MBP structure. Dashed lines and the corresponding coloured symbols correspond to the whole FOV velocities (blue) and all MBPs (green). The latter two distributions were rescaled to the number of tracks, i.e., number of measurements contained in the first three distributions. The blue curve is centred on the zero velocity level by the calibration of the inversion data (setting the mean quiet Sun flows on average to zero).
\label{figure14}}
\end{figure}

\begin{figure}
\begin{center}
\includegraphics[width=0.45\textwidth]{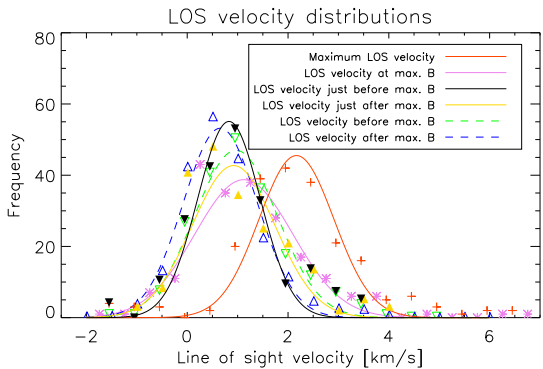}
\end{center}
\caption{Line-of-sight velocity distributions of MBPs in different evolutionary stages, similar to Fig. \ref{figure14}. Here the distributions of LOS velocities before (green open downside pointing triangles) and after (blue open upside pointing triangles) the magnetic field strength maximum is reached (the distributions for the ascending and descending phase of the magnetic field evolution) are shown together with the distributions at the magnetic field strength maximum (violet stars) as well as prior and posterior to this moment (black solid downward-pointing triangles and yellow solid upward-pointing triangles), and finally the distribution comprising the maximum LOS velocities measured during each of the evolutions of the MBPs (red crosses). The distributions for the ascending and descending phases have been normalised to the number of tracks for comparison reasons.
\label{figure14a}}
\end{figure}

\subsection{Evolutionary MBP track statistics:}
An important part of the current study considered the statistics of evolutionary MBP tracks and the results gained by such statistics. 

Let us now discuss in more detail the statistics of the magnetic field strength evolution. As we found in Sect. 4.4, the distribution of initial magnetic field strengths is centred around the equipartition field strength (actually the mean value of all the initial fields is 0.46 $\pm$ 0.17 kG). MBPs are clearly distinct from the background magnetic field. This can be seen in Fig. \ref{figure13} too. Here we plot the distribution of the initial magnetic field strengths (yellow triangles), the maximum magnetic field strength reached during the evolution (orange crosses), the dissolution field strength (black asterisks; average value of 0.41 $\pm$ 0.18 kG), and the background magnetic field strength of the complete FOV (blue diamonds). The complete magnetic field strength distribution, considering all single measurements (i.e., every barycentre magnetic field strength of the identified MBPs in all frames), is represented by green circles. The correspondingly coloured lines are the results of best fits of log-normal distributions to the data. Such log-normal distributions were already found for other characteristic MBP parameters in the past \citep[e.g., for the feature size;][]{2010ApJ...725L.101A,2010ApJ...722L.188C} and can be related to the fragmentation processes MBPs are undergoing \citep[splitting and merging; see, e.g.,][]{1988ApJ...327..451B,Eckhard}. The distributions turned out to be fair representations of the distributions except for the maximum field strengths that tend to show a bimodal distribution with values either below or above 1 kG. Such bi-modality was already reported in \citet{2013A&A...554A..65U}, who speculated that the two maxima might be related to internetwork fields and network fields. 

This coincides with results of \citet{2007A&A...472..607B} who investigated the magnetic field strength distribution (and other parameters) of MBPs around a sunspot moat and found a continuous distribution of field strengths starting with values around 0.5 kG up to 1.5 kG. On the other hand \citet{2010ApJ...723L.164L} showed (using the same data sets as in the current study) that at least some of the magnetic features in the internetwork have kG strength. Also, bear in mind the work of \citet[][]{2014A&A...568A..13R} who show that only magnetic features with kG fields are associated with a clear MBP, but that many of these appear as weaker fields in the observations, due to the insufficient spatial resolution. Furthermore straylight might affect the Stokes measurements and lead to a reduced detected magnetic field strength. A similar conclusion was also drawn by \citet{2014A&A...562L...1C}, who used simulated data and investigated the behaviour of the field strength distribution with changing spatial resolutions and instrument effects. 

The dashed lines (FOV distribution and distribution of all MBP measurements) were normalised to the same total frequency (number of measurements) as for the other distributions, i.e., to the number of analysed MBP tracks. This was done to allow a better comparison between the distributions. Else the FOV distribution and the distribution of all MBP measurements would be on a different absolute scale (far away from the others). The vertical dash-dotted lines in Fig. \ref{figure13} indicate the average background noise magnetic field strength over the FOV (0.14 kG; see Sect. 4.4) and an upper boundary of the equipartition field (0.4 kG; see \citealt{1996A&A...310L..33S}, who gives a range for the equipartition field from 0.2 to 0.4 kG). Note that the mean value indeed coincides with the expectation value of the (log-normal) distribution of magnetic field strengths over the full FOV. Such an expectation value is given for a log-normal distribution in general by 
\begin{equation}
 E(X)=\exp\left(\mu+\frac{\sigma^2}{2}\right),
 \end{equation}
  where $\mu$ and $\sigma$ are the mean and standard deviation of the natural logarithm of $X$. 
We see that the initial magnetic field strength distribution is separated by two times the rms value from the distribution of the complete FOV. We stated before that the expectation value (generally also known as mean value) for a log-normal distribution is situated at the right slope of the distribution due to its skewness. Very interestingly, the expectation value of the initial magnetic field strength distribution is 0.424 kG, which lies within the range of reasonable values for the equipartition field strength.

This discussion supports the previously stated result that MBPs, which resemble the strongest part of magnetic structures (the core of magnetic structures) seem to be formed out of larger patches of equipartition magnetic fields and possess a distinct magnetic field strength distribution that is distinct from the background magnetic fields. This result is in agreement with the suggestion made by \citet{1986Natur.322..156V} that for the convective collapse to be successful the magnetic features must have a minimum size, i.e., a minimum amount of magnetic flux \citep[as was observationally confirmed by][]{1996A&A...310L..33S}. Indeed, 30 \% (63 out 207) tracked MBPs reach kG field strengths at a given moment of their lifetime (although this fraction may be influenced by the limited spatial resolution).


\begin{figure}
\begin{center}
\includegraphics[width=0.45\textwidth]{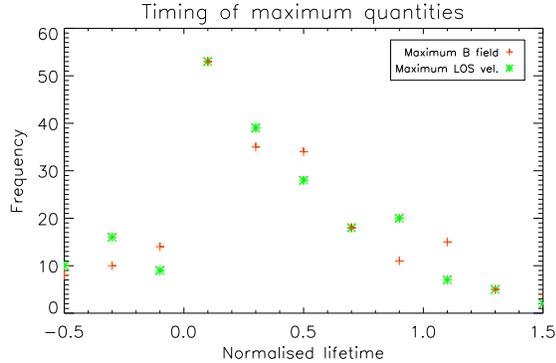}
\end{center}
\caption{Timing of the maximum in the magnetic field strength and the LOS velocity. The frequency of the maxima of these quantities vs. the normalised MBP lifetime is shown. Red crosses illustrate the histogram of the times at which the maximum field strength is reached and green asterisks times of maximum LOS velocity.
\label{figure15a}}
\end{figure}

As we have seen in Sect. 4.4 and in Fig. \ref{figure11}, initial downflows are on the order of 1 to 2 km/s, which increase two- to threefold at some point during their evolution, reaching values from 4 to 6 km/s. We show in Fig. \ref{figure14} the LOS velocity distributions of the initial measurement (yellow triangles), the distribution of maximum LOS velocities (orange crosses) reached during the evolution and the dissolution distribution (black asterisks). Furthermore, as before, we depict the distribution of all LOS velocities in the FOV for the sake of comparison (blue diamonds) and of all single MBP measurements (green circles). The latter two distributions were rescaled to the number of tracks (corresponding to the measurements entering the previous three distributions) to enable an easier comparison. The lines are best-fit Gaussians. Although the different fits may not be equally good for all the distributions, their Gaussian mean parameters can easily be used to see that MBPs statistically start to be detected when downflows have already developed (0.8 km/s), that maximum velocities are clearly higher (2.4 km/s), and that dynamic effects fade away at the same time that the MBPs disappear (0.6 km/s). Interestingly, the green curve shows a maximum very close to that of the orange curve, i.e., below 1 km/s. The main difference is that the green curve is somewhat broader, with larger, but also a few smaller velocities. This suggests that for most of their time, the MBPs have velocities similar to that at the beginning.

To study in more detail the pre- and post-collapse phase we investigate the distribution of LOS velocities prior and posterior to the moment that the maximum field strength is reached, let us call this time $t_{\rm max}$. Figure \ref{figure14a} shows the histogram of the LOS velocities prior to  $t_{\rm max}$ as black solid line (downward-pointing solid triangles) and posterior to $t_{\rm max}$ as yellow line (solid upward-pointing triangles). Moreover the green dashed curve gives the ascending phase of the MBP evolution and the descending phase of the evolution after $t_{\rm max}$ is shown in blue colour. The latter four curves are slightly shifted relative to each other suggesting that the velocities prior to the maximum field strength are slightly higher compared to the posterior phase. Additionally the LOS distributions just prior and posterior to $t_{\rm max}$ seem to be more skewed and hence also deviate more from Gaussian fits (illustrated by the corresponding lines) than the distributions for the full ascending/descending phase. The distribution at $t_{\rm max}$ is plotted in violet colour and for comparison we display again the distribution of maximum LOS velocities during the features' lifetime. Interestingly the distribution at $t_{\rm max}$ is clearly shifted to lower values compared to the maximum LOS velocity distribution (red colour) suggesting that the highest velocities during the MBP evolution are not reached at exactly $t_{\rm max}$. Nevertheless the velocities during the maximum are on average higher than at the other phases of the evolution (green/blue curve). The green and the blue curve have been again normalised to the number of tracks, as in Fig. \ref{figure14}. Special interest is often drawn to the question of flow reversals or upflows during the evolution of the MBP. An investigation of the found distribution yields values of 13 \% of upflows prior to $t_{\rm max}$ and 14\% just posterior to this moment, which is therefore not significantly different. Interestingly this picture changes if we have a look at the distribution right at $t_{\rm max}$. Here we can only find in 10 \% of the cases an upflow. Additionally the evolution seems to be not symmetric, which we can see if we consider all the measurements during the two phases of the MBP evolution (the phase prior to $t_{\rm max}$ and posterior). Here we find only 11 \% of the LOS measurements giving upflows in the ascending phase of the magnetic field evolution but 17 \% of the LOS measurements give upflows in the descending phase of the magnetic field. Therefore it is probably safe to say that even if we cannot find direct flow reversals and strong upflows after the creation of the MBPs, the phase after $t_{\rm max}$ has generally speaking weaker downflows and exhibits also more often some upflows.

The last detail we focus on is the relative timing of the maximum seen in different plasma quantities of the MBP. The lifetime of the MBPs has been normalised for that purpose. Negative values mean in that respect that the maximum occurred already before the tracking and a value greater of one that the maximum happened in the post-tracking phase. 
The resulting histograms of the maximum field strength (marked by red crosses) and the maximum LOS velocity (green asterisks are plotted in Fig. \ref{figure15a}). 
Interestingly both quantities show their maximum just in the beginning of the evolution. The coincidence in the timing of the maxima of both quantities can be considered in agreement with former studies having less spatial and temporal resolution. It is not at odds with the previous finding of asynchrony between field strength and LOS velocity maxima in some specific cases since the histogram only gives statistical information. It does not say that for a particular feature the maximum velocity must occur at the same time as the maximum field strength. 

\section{Summary and conclusions}
In this study we tracked the evolution of small-scale magnetic fields that manifest themselves as MBPs in spectropolarimetric data observed with \textit{Sunrise}/IMaX. The used data set was taken during a period of exceptionally quiet Sun close to the solar disc centre in 2009 June. The identification and tracking of the features were done automatically in the blue Fe line wing ($-$40 m\AA~off the line centre) which appears by visual inspection to have similar properties to the G-band for the identification of MBPs although this proxy has to be used carefully as the intensity might be influenced by vertical motions along the LOS. Finally, plasma parameters of interest were extracted from fully processed maps obtained by the inversion code SIR.

The three case studies of MBP evolution presented in detail illustrate the variety of evolutionary histories and indicate that a variety of processes may be involved in the creation of MBPs as well as for stabilising them during their lifetime. In fact the observed plasma parameters are in a complicated way related to each other and drawing conclusions from one single observed track would not be generally applicable and may be misleading. One conclusion which can be drawn, however, is that MBPs, at least the ones in the quiet Sun, are rapidly evolving structures with constant, significant fluctuations, i.e., most of them are transient features and the longer lived ones show several brightness and field strength peaks. A similar conclusion was drawn by \citet{2013JPhCS.440a2032J}, \citet{2011A&A...529A..79N}, and by \citet{2009A&A...504..583F}, who did a statistical analysis of convective collapse events in a \textit{Hinode} data set.

We have in addition obtained more generally applicable results by considering all the observed tracks:
\begin{itemize}
 \item The \textit{magnetic field strength} starts weak with values near the upper end of estimates of the equipartition field strength of about 0.4 kG, reaches maximum values two to three times the initial value, which is in most cases still below 1 kG probably also caused by resolution effects, and declines to values consistent with the equipartition field strength before the feature dissolves.
\item The \textit{LOS velocity} shows a downflow practically all the time. It starts with a weak value of about 0.8 km/s, reaches a clear maximum (in a few cases of up to 6 km/s but on average 2.4 km/s) before declining again to about 0.6 km/s on average, i.e., a value smaller than the starting value.
\end{itemize}

These observational facts are partly congruent with the expectations one would have from the convective collapse scenario and confirm on a statistical basis some of the expected implications of the theory such as an increase of the magnetic field strength coupled with strong, transient downflows. The presence of significant downflows co-located with the bright point throughout its lifetime of most of our features is in complete contrast with the absence of such downflows in bright points in active region plage. 
Moreover, we have outlined in the second studied case the importance of near small-scale upflows. 
The importance, evolution, and driving force of such spatially co-located up and downflows (shear flows) have to be studied in more detail in the future. Another point to be investigated in the future is the decay of the features. Are MBPs really disintegrated by a downflow reversal leading to an upwards drifting shock front? If so, then why are upflows recorded so rarely in our statistical sample? Only 33 out of a total of 207 investigated MBPs displayed upflows just before they disappeared as bright points. An even smaller fraction of MBPs display upflows just before or at the time their B values start to decrease. These values suggest that another mechanism is responsible for the weakening of the field than strong upflows. Naturally, we cannot exclude the possibility that upflows \citep[maybe caused by upwards propagating shocks as discussed in][]{2001ApJ...560.1010B} are too quickly evolving so that they cannot be detected within the temporal resolution of about 33 s at hand. Some of the previous questions have been partially addressed by \citet{2014ApJ...789....6R}, but it would be worthwhile to search for longer and more stable MBP evolutions \citep[such as the one reported by][]{2008ApJ...677L.145N} and to investigate what physical processes lead to the stabilisation. Due to their lateral expansion with height, the surface of flux tubes is in principle susceptible to the hydromagnetic exchange instability \citep[e.g.,][]{1975SoPh...40..291P}. Therefore several possibilities were proposed in literature how to overcome the instabilities and stabilise flux tubes. This could be achieved by suitable environmental conditions such as the right temperature fields \citep{1993A&A...268..299B,1993A&A...273..287B} or swirl flows \citep[][]{1984A&A...140..453S} around the magnetic flux tube. The latter is definitely worthwhile to be studied in more detail as recently the interest in swirls and vortices \citep[e.g.][]{2008ApJ...687L.131B,2011A&A...526A...5S} in the solar atmosphere has gained a lot of interest again due to new observations showing clearly such ``tornado-like'' structures \citep[e.g.][]{2012Natur.486..505W,2013JPhCS.440a2005W}.

Another rather interesting follow-up study would be to check if the downflows (and hence the MBPs) studied here were preceded by magnetic flux emergence (e.g. as seen in linear polarization), i.e., could these downflows really be the signature of convective collapse, or rather are they simply the draining of rising loops that emerged recently? Possibly there is a component of both in the observed downflows associated with the enhancement of the magnetic field strength?

\acknowledgements
The research was funded by the Austrian Science Fund (FWF): J3176. In addition D.U. wishes to thank the {\"O}sterreichischer Austauschdienst ({\"O}AD) and the Ministry of Education Youth and Sports (M\v{S}MT) of the Czech Republic for financing a
short research stay at the Astronomical Institute of the Czech Academy of
Sciences in Ondrejov in the frame of the project MEB061109. Furthermore, J.J. wants to express vice versa his gratitude to the M\v{S}MT and {\"O}AD for financing a short research stay at the IGAM of the University of Graz. Moreover, J.J. is grateful for support from the Czech Science Foundation (GACR) through project~P209/12/0287 and RVO:67985815. Partial funding has also been obtained from the Spanish Ministerio de Econom\'{i}a through Projects AYA2011-29833-C06 and AYA2012-39636-C06, including a percentage of European FEDER funds. The German contribution to \textit{Sunrise} is funded by the Bundesministerium f\"ur Wirtschaft und Technologie through Deutsches Zentrum f\"ur Luft- und Raumfahrt e.V. (DLR), Grant No. 50OU 0401, and by the Innovationsfond of the President of the Max Planck Society (MPG). The High Altitude Observatory (HAO) contribution was partly funded through NASA grant NNX08AH38G. This work was partly supported by the BK21 plus program through the National Research Foundation (NRF) funded by the Ministry of Education of Korea. Finally the authors wish to express their sincere gratitude to the anonymous referee whose comments and remarks helped significantly to improve the manuscript.
Moreover, the whole department is grateful for the English proof reader for creating a heated debate about the possibility for introducing new words into the English language. To put it into the words of the proof reader: ``The words ``till'' and ``still'' have been changed to ``suntil'' here and elsewhere in the text. Please confirm if it
is correct.''





\bibliographystyle{apj}
\bibliography{apj-jour,Literaturverzeichnis}

\begin{thebibliography}{93}
\expandafter\ifx\csname natexlab\endcsname\relax\def\natexlab#1{#1}\fi

\bibitem[{{Abramenko} {et~al.}(2010){Abramenko}, {Yurchyshyn}, {Goode}, \&
  {Kilcik}}]{2010ApJ...725L.101A}
{Abramenko}, V., {Yurchyshyn}, V., {Goode}, P., \& {Kilcik}, A. 2010, \apjl,
  725, L101

\bibitem[{{Abramenko} {et~al.}(2012){Abramenko}, {Yurchyshyn}, {Goode},
  {Kitiashvili}, \& {Kosovichev}}]{2012ApJ...756L..27A}
{Abramenko}, V.~I., {Yurchyshyn}, V.~B., {Goode}, P.~R., {Kitiashvili}, I.~N.,
  \& {Kosovichev}, A.~G. 2012, \apjl, 756, L27

\bibitem[{{Andi{\'c}} {et~al.}(2011){Andi{\'c}}, {Chae}, {Goode}, {Cao}, {Ahn},
  {Yurchyshyn}, \& {Abramenko}}]{2011ApJ...731...29A}
{Andi{\'c}}, A., {Chae}, J., {Goode}, P.~R., {et~al.} 2011, \apj, 731, 29

\bibitem[{{Barthol} {et~al.}(2011){Barthol}, {Gandorfer}, {Solanki},
  {Sch{\"u}ssler}, {Chares}, {Curdt}, {Deutsch}, {Feller}, {Germerott},
  {Grauf}, {Heerlein}, {Hirzberger}, {Kolleck}, {Meller}, {M{\"u}ller},
  {Riethm{\"u}ller}, {Tomasch}, {Kn{\"o}lker}, {Lites}, {Card}, {Elmore},
  {Fox}, {Lecinski}, {Nelson}, {Summers}, {Watt}, {Mart{\'{\i}}nez Pillet},
  {Bonet}, {Schmidt}, {Berkefeld}, {Title}, {Domingo}, {Gasent Blesa}, {Del
  Toro Iniesta}, {L{\'o}pez Jim{\'e}nez}, {{\'A}lvarez-Herrero},
  {Sabau-Graziati}, {Widani}, {Haberler}, {H{\"a}rtel}, {Kampf}, {Levin},
  {P{\'e}rez Grande}, {Sanz-Andr{\'e}s}, \& {Schmidt}}]{2011SoPh..268....1B}
{Barthol}, P., {Gandorfer}, A., {Solanki}, S.~K., {et~al.} 2011, \solphys, 268,
  1

\bibitem[{{Beck} {et~al.}(2007){Beck}, {Bellot Rubio}, {Schlichenmaier}, \&
  {S{\"u}tterlin}}]{2007A&A...472..607B}
{Beck}, C., {Bellot Rubio}, L.~R., {Schlichenmaier}, R., \& {S{\"u}tterlin}, P.
  2007, \aap, 472, 607

\bibitem[{{Bellot Rubio} {et~al.}(2001){Bellot Rubio}, {Rodr{\'{\i}}guez
  Hidalgo}, {Collados}, {Khomenko}, \& {Ruiz Cobo}}]{2001ApJ...560.1010B}
{Bellot Rubio}, L.~R., {Rodr{\'{\i}}guez Hidalgo}, I., {Collados}, M.,
  {Khomenko}, E., \& {Ruiz Cobo}, B. 2001, \apj, 560, 1010

\bibitem[{{Berger} {et~al.}(1998){Berger}, {L{\"o}fdahl}, {Shine}, \&
  {Title}}]{1998ApJ...506..439B}
{Berger}, T.~E., {L{\"o}fdahl}, M.~G., {Shine}, R.~A., \& {Title}, A.~M. 1998,
  \apj, 506, 439

\bibitem[{{Berger} {et~al.}(1995){Berger}, {Schrijver}, {Shine}, {Tarbell},
  {Title}, \& {Scharmer}}]{1995ApJ...454..531B}
{Berger}, T.~E., {Schrijver}, C.~J., {Shine}, R.~A., {et~al.} 1995, \apj, 454,
  531

\bibitem[{{Berger} \& {Title}(1996)}]{1996ApJ...463..365B}
{Berger}, T.~E., \& {Title}, A.~M. 1996, \apj, 463, 365

\bibitem[{{Berger} {et~al.}(2004){Berger}, {Rouppe van der Voort},
  {L{\"o}fdahl}, {Carlsson}, {Fossum}, {Hansteen}, {Marthinussen}, {Title}, \&
  {Scharmer}}]{2004A&A...428..613B}
{Berger}, T.~E., {Rouppe van der Voort}, L.~H.~M., {L{\"o}fdahl}, M.~G.,
  {et~al.} 2004, \aap, 428, 613

\bibitem[{{Berkefeld} {et~al.}(2011){Berkefeld}, {Schmidt}, {Soltau}, {Bell},
  {Doerr}, {Feger}, {Friedlein}, {Gerber}, {Heidecke}, {Kentischer},
  {v.~D.~L{\"u}he}, {Sigwarth}, {W{\"a}lde}, {Barthol}, {Deutsch}, {Gandorfer},
  {Germerott}, {Grauf}, {Meller}, {{\'A}lvarez-Herrero}, {Kn{\"o}lker},
  {Mart{\'{\i}}nez Pillet}, {Solanki}, \& {Title}}]{2011SoPh..268..103B}
{Berkefeld}, T., {Schmidt}, W., {Soltau}, D., {et~al.} 2011, \solphys, 268, 103

\bibitem[{{Bogdan} {et~al.}(1988){Bogdan}, {Gilman}, {Lerche}, \&
  {Howard}}]{1988ApJ...327..451B}
{Bogdan}, T.~J., {Gilman}, P.~A., {Lerche}, I., \& {Howard}, R. 1988, \apj,
  327, 451

\bibitem[{{Bonet} {et~al.}(2008){Bonet}, {M{\'a}rquez}, {S{\'a}nchez Almeida},
  {Cabello}, \& {Domingo}}]{2008ApJ...687L.131B}
{Bonet}, J.~A., {M{\'a}rquez}, I., {S{\'a}nchez Almeida}, J., {Cabello}, I., \&
  {Domingo}, V. 2008, \apjl, 687, L131

\bibitem[{{Borrero} \& {Kobel}(2011)}]{2011A&A...527A..29B}
{Borrero}, J.~M., \& {Kobel}, P. 2011, \aap, 527, A29

\bibitem[{{Borrero} \& {Kobel}(2012)}]{2012A&A...547A..89B}
---. 2012, \aap, 547, A89

\bibitem[{{Bovelet} \& {Wiehr}(2008)}]{2008A&A...488.1101B}
{Bovelet}, B., \& {Wiehr}, E. 2008, \aap, 488, 1101

\bibitem[{{B\"uhler}(2013)}]{Buehler}
{B\"uhler}, D. 2013, PhD thesis, PROPHYS

\bibitem[{{B\"unte} {et~al.}(1993{\natexlab{a}}){B\"unte}, {Hasan}, \&
  {Kalkofen}}]{1993A&A...273..287B}
{B\"unte}, M., {Hasan}, S., \& {Kalkofen}, W. 1993{\natexlab{a}}, \aap, 273,
  287

\bibitem[{{B\"unte} {et~al.}(1993{\natexlab{b}}){B\"unte}, {Steiner}, \&
  {Pizzo}}]{1993A&A...268..299B}
{B\"unte}, M., {Steiner}, O., \& {Pizzo}, V.~J. 1993{\natexlab{b}}, \aap, 268,
  299

\bibitem[{{Chitta} {et~al.}(2012){Chitta}, {van Ballegooijen}, {Rouppe van der
  Voort}, {DeLuca}, \& {Kariyappa}}]{2012ApJ...752...48C}
{Chitta}, L.~P., {van Ballegooijen}, A.~A., {Rouppe van der Voort}, L.,
  {DeLuca}, E.~E., \& {Kariyappa}, R. 2012, \apj, 752, 48

\bibitem[{{Criscuoli} \& {Uitenbroek}(2014)}]{2014A&A...562L...1C}
{Criscuoli}, S., \& {Uitenbroek}, H. 2014, \aap, 562, L1

\bibitem[{{Crockett} {et~al.}(2010){Crockett}, {Mathioudakis}, {Jess},
  {Shelyag}, {Keenan}, \& {Christian}}]{2010ApJ...722L.188C}
{Crockett}, P.~J., {Mathioudakis}, M., {Jess}, D.~B., {et~al.} 2010, \apjl,
  722, L188

\bibitem[{{Danilovic} {et~al.}(2010){Danilovic}, {Sch{\"u}ssler}, \&
  {Solanki}}]{2010A&A...509A..76D}
{Danilovic}, S., {Sch{\"u}ssler}, M., \& {Solanki}, S.~K. 2010, \aap, 509, A76

\bibitem[{{de Wijn} {et~al.}(2005){de Wijn}, {Rutten}, {Haverkamp}, \&
  {S{\"u}tterlin}}]{2005AA...441.1183D}
{de Wijn}, A.~G., {Rutten}, R.~J., {Haverkamp}, E.~M.~W.~P., \&
  {S{\"u}tterlin}, P. 2005, \aap, 441, 1183

\bibitem[{{de Wijn} {et~al.}(2009){de Wijn}, {Stenflo}, {Solanki}, \&
  {Tsuneta}}]{2009SSRv..144..275D}
{de Wijn}, A.~G., {Stenflo}, J.~O., {Solanki}, S.~K., \& {Tsuneta}, S. 2009,
  \ssr, 144, 275

\bibitem[{{DeForest} {et~al.}(2007){DeForest}, {Hagenaar}, {Lamb}, {Parnell},
  \& {Welsch}}]{2007ApJ...666..576D}
{DeForest}, C.~E., {Hagenaar}, H.~J., {Lamb}, D.~A., {Parnell}, C.~E., \&
  {Welsch}, B.~T. 2007, \apj, 666, 576

\bibitem[{{Dunn} \& {Zirker}(1973)}]{1973SoPh...33..281D}
{Dunn}, R.~B., \& {Zirker}, J.~B. 1973, \solphys, 33, 281

\bibitem[{{Fischer} {et~al.}(2009){Fischer}, {de Wijn}, {Centeno}, {Lites}, \&
  {Keller}}]{2009A&A...504..583F}
{Fischer}, C.~E., {de Wijn}, A.~G., {Centeno}, R., {Lites}, B.~W., \& {Keller},
  C.~U. 2009, \aap, 504, 583

\bibitem[{{Gandorfer} {et~al.}(2011){Gandorfer}, {Grauf}, {Barthol},
  {Riethm{\"u}ller}, {Solanki}, {Chares}, {Deutsch}, {Ebert}, {Feller},
  {Germerott}, {Heerlein}, {Heinrichs}, {Hirche}, {Hirzberger}, {Kolleck},
  {Meller}, {M{\"u}ller}, {Sch{\"a}fer}, {Tomasch}, {Kn{\"o}lker},
  {Mart{\'{\i}}nez Pillet}, {Bonet}, {Schmidt}, {Berkefeld}, {Feger},
  {Heidecke}, {Soltau}, {Tischenberg}, {Fischer}, {Title}, {Anwand}, \&
  {Schmidt}}]{2011SoPh..268...35G}
{Gandorfer}, A., {Grauf}, B., {Barthol}, P., {et~al.} 2011, \solphys, 268, 35

\bibitem[{{Giannattasio} {et~al.}(2013){Giannattasio}, {Del Moro}, {Berrilli},
  {Bellot Rubio}, {Gos{\#728}i{\'c}}, \& {Orozco
  Su{\'a}rez}}]{2013ApJ...770L..36G}
{Giannattasio}, F., {Del Moro}, D., {Berrilli}, F., {et~al.} 2013, \apjl, 770,
  L36

\bibitem[{{Grossmann-Doerth} {et~al.}(1998){Grossmann-Doerth}, {Sch\"ussler},
  \& {Steiner}}]{1998A&A...337..928G}
{Grossmann-Doerth}, U., {Sch\"ussler}, M., \& {Steiner}, O. 1998, \aap, 337,
  928

\bibitem[{{Hagenaar} {et~al.}(2003){Hagenaar}, {Schrijver}, \&
  {Title}}]{2003ApJ...584.1107H}
{Hagenaar}, H.~J., {Schrijver}, C.~J., \& {Title}, A.~M. 2003, \apj, 584, 1107

\bibitem[{{Hagenaar} {et~al.}(1999){Hagenaar}, {Schrijver}, {Title}, \&
  {Shine}}]{1999ApJ...511..932H}
{Hagenaar}, H.~J., {Schrijver}, C.~J., {Title}, A.~M., \& {Shine}, R.~A. 1999,
  \apj, 511, 932

\bibitem[{{Hewitt} {et~al.}(2014){Hewitt}, {Shelyag}, {Mathioudakis}, \&
  {Keenan}}]{2014A&A...565A..84H}
{Hewitt}, R.~L., {Shelyag}, S., {Mathioudakis}, M., \& {Keenan}, F.~P. 2014,
  \aap, 565, A84

\bibitem[{{Jafarzadeh} {et~al.}(2013){Jafarzadeh}, {Solanki}, {Feller}, {Lagg},
  {Pietarila}, {Danilovic}, {Riethm{\"u}ller}, \& {Mart{\'{\i}}nez
  Pillet}}]{2013A&A...549A.116J}
{Jafarzadeh}, S., {Solanki}, S.~K., {Feller}, A., {et~al.} 2013, \aap, 549,
  A116

\bibitem[{{Joshi} {et~al.}(2011){Joshi}, {Pietarila}, {Hirzberger}, {Solanki},
  {Aznar Cuadrado}, \& {Merenda}}]{2011ApJ...734L..18J}
{Joshi}, J., {Pietarila}, A., {Hirzberger}, J., {et~al.} 2011, \apjl, 734, L18

\bibitem[{{Jur{\v c}{\'a}k} {et~al.}(2013){Jur{\v c}{\'a}k}, {Utz}, \& {Bellot
  Rubio}}]{2013JPhCS.440a2032J}
{Jur{\v c}{\'a}k}, J., {Utz}, D., \& {Bellot Rubio}, L.~R. 2013, Journal of
  Physics Conference Series, 440, 012032

\bibitem[{{Kato} {et~al.}(2011){Kato}, {Steiner}, {Steffen}, \&
  {Suematsu}}]{2011ApJ...730L..24K}
{Kato}, Y., {Steiner}, O., {Steffen}, M., \& {Suematsu}, Y. 2011, \apjl, 730,
  L24

\bibitem[{{Keller} {et~al.}(2004){Keller}, {Sch{\"u}ssler}, {V{\"o}gler}, \&
  {Zakharov}}]{2004ApJ...607L..59K}
{Keller}, C.~U., {Sch{\"u}ssler}, M., {V{\"o}gler}, A., \& {Zakharov}, V. 2004,
  \apjl, 607, L59

\bibitem[{{Keys} {et~al.}(2011){Keys}, {Mathioudakis}, {Jess}, {Shelyag},
  {Crockett}, {Christian}, \& {Keenan}}]{2011ApJ...740L..40K}
{Keys}, P.~H., {Mathioudakis}, M., {Jess}, D.~B., {et~al.} 2011, \apjl, 740,
  L40

\bibitem[{{Khomenko} {et~al.}(2003){Khomenko}, {Collados}, {Solanki}, {Lagg},
  \& {Trujillo Bueno}}]{2003A&A...408.1115K}
{Khomenko}, E.~V., {Collados}, M., {Solanki}, S.~K., {Lagg}, A., \& {Trujillo
  Bueno}, J. 2003, \aap, 408, 1115

\bibitem[{{Klimchuk}(2006)}]{2006SoPh..234...41K}
{Klimchuk}, J.~A. 2006, \solphys, 234, 41

\bibitem[{{Lagg} {et~al.}(2010){Lagg}, {Solanki}, {Riethm{\"u}ller},
  {Mart{\'{\i}}nez Pillet}, {Sch{\"u}ssler}, {Hirzberger}, {Feller}, {Borrero},
  {Schmidt}, {del Toro Iniesta}, {Bonet}, {Barthol}, {Berkefeld}, {Domingo},
  {Gandorfer}, {Kn{\"o}lker}, \& {Title}}]{2010ApJ...723L.164L}
{Lagg}, A., {Solanki}, S.~K., {Riethm{\"u}ller}, T.~L., {et~al.} 2010, \apjl,
  723, L164

\bibitem[{{Langangen} {et~al.}(2007){Langangen}, {Carlsson}, {Rouppe van der
  Voort}, \& {Stein}}]{2007ApJ...655..615L}
{Langangen}, {\O}., {Carlsson}, M., {Rouppe van der Voort}, L., \& {Stein},
  R.~F. 2007, \apj, 655, 615

\bibitem[{{Lemmerer} {et~al.}(2014){Lemmerer}, {Utz}, {Hanslmeier}, {Veronig},
  {Thonhofer}, {Grimm-Strele}, \& {Kariyappa}}]{2014A&A...563A.107L}
{Lemmerer}, B., {Utz}, D., {Hanslmeier}, A., {et~al.} 2014, \aap, 563, A107

\bibitem[{{Limpert} {et~al.}(2008){Limpert}, {Stahel}, \& {Abbt}}]{Eckhard}
{Limpert}, E., {Stahel}, W.~A., \& {Abbt}, M. 2008, BioScience, 51, 341

\bibitem[{{Lin} \& {Rimmele}(1999)}]{1999ApJ...514..448L}
{Lin}, H., \& {Rimmele}, T. 1999, \apj, 514, 448

\bibitem[{{Mart{\'{\i}}nez Pillet} {et~al.}(2011){Mart{\'{\i}}nez Pillet}, {Del
  Toro Iniesta}, {{\'A}lvarez-Herrero}, {Domingo}, {Bonet}, {Gonz{\'a}lez
  Fern{\'a}ndez}, {L{\'o}pez Jim{\'e}nez}, {Pastor}, {Gasent Blesa}, {Mellado},
  {Piqueras}, {Aparicio}, {Balaguer}, {Ballesteros}, {Belenguer}, {Bellot
  Rubio}, {Berkefeld}, {Collados}, {Deutsch}, {Feller}, {Girela}, {Grauf},
  {Heredero}, {Herranz}, {Jer{\'o}nimo}, {Laguna}, {Meller}, {Men{\'e}ndez},
  {Morales}, {Orozco Su{\'a}rez}, {Ramos}, {Reina}, {Ramos},
  {Rodr{\'{\i}}guez}, {S{\'a}nchez}, {Uribe-Patarroyo}, {Barthol}, {Gandorfer},
  {Knoelker}, {Schmidt}, {Solanki}, \& {Vargas
  Dom{\'{\i}}nguez}}]{2011SoPh..268...57M}
{Mart{\'{\i}}nez Pillet}, V., {Del Toro Iniesta}, J.~C., {{\'A}lvarez-Herrero},
  A., {et~al.} 2011, \solphys, 268, 57

\bibitem[{{Muller}(1983)}]{1983SoPh...85..113M}
{Muller}, R. 1983, \solphys, 85, 113

\bibitem[{{Muller} {et~al.}(1989){Muller}, {Hulot}, \&
  {Roudier}}]{1989SoPh..119..229M}
{Muller}, R., {Hulot}, J.~C., \& {Roudier}, T. 1989, \solphys, 119, 229

\bibitem[{{Muller} {et~al.}(2011){Muller}, {Utz}, \&
  {Hanslmeier}}]{2011SoPh..274...87M}
{Muller}, R., {Utz}, D., \& {Hanslmeier}, A. 2011, \solphys, 274, 87

\bibitem[{{Nagata} {et~al.}(2008){Nagata}, {Tsuneta}, {Suematsu}, {Ichimoto},
  {Katsukawa}, {Shimizu}, {Yokoyama}, {Tarbell}, {Lites}, {Shine}, {Berger},
  {Title}, {Bellot Rubio}, \& {Orozco Su{\'a}rez}}]{2008ApJ...677L.145N}
{Nagata}, S., {Tsuneta}, S., {Suematsu}, Y., {et~al.} 2008, \apjl, 677, L145

\bibitem[{{Narayan}(2011)}]{2011A&A...529A..79N}
{Narayan}, G. 2011, \aap, 529, A79

\bibitem[{{Narayan} \& {Scharmer}(2010)}]{2010A&A...524A...3N}
{Narayan}, G., \& {Scharmer}, G.~B. 2010, \aap, 524, A3

\bibitem[{{Nisenson} {et~al.}(2003){Nisenson}, {van Ballegooijen}, {de Wijn},
  \& {S{\"u}tterlin}}]{2003ApJ...587..458N}
{Nisenson}, P., {van Ballegooijen}, A.~A., {de Wijn}, A.~G., \&
  {S{\"u}tterlin}, P. 2003, \apj, 587, 458

\bibitem[{{Parker}(1975)}]{1975SoPh...40..291P}
{Parker}, E.~N. 1975, \solphys, 40, 291

\bibitem[{{Parker}(1978)}]{1978ApJ...221..368P}
---. 1978, \apj, 221, 368

\bibitem[{{Requerey} {et~al.}(2014){Requerey}, {Del Toro Iniesta}, {Bellot
  Rubio}, {Bonet}, {Mart{\'{\i}}nez Pillet}, {Solanki}, \&
  {Schmidt}}]{2014ApJ...789....6R}
{Requerey}, I.~S., {Del Toro Iniesta}, J.~C., {Bellot Rubio}, L.~R., {et~al.}
  2014, \apj, 789, 6

\bibitem[{{Riethm{\"u}ller} {et~al.}(2014){Riethm{\"u}ller}, {Solanki},
  {Berdyugina}, {Sch{\"u}ssler}, {Mart{\'{\i}}nez Pillet}, {Feller},
  {Gandorfer}, \& {Hirzberger}}]{2014A&A...568A..13R}
{Riethm{\"u}ller}, T.~L., {Solanki}, S.~K., {Berdyugina}, S.~V., {et~al.} 2014,
  \aap, 568, A13

\bibitem[{{Riethm{\"u}ller} {et~al.}(2010){Riethm{\"u}ller}, {Solanki},
  {Mart{\'{\i}}nez Pillet}, {Hirzberger}, {Feller}, {Bonet}, {Bello
  Gonz{\'a}lez}, {Franz}, {Sch{\"u}ssler}, {Barthol}, {Berkefeld}, {del Toro
  Iniesta}, {Domingo}, {Gandorfer}, {Kn{\"o}lker}, \&
  {Schmidt}}]{2010ApJ...723L.169R}
{Riethm{\"u}ller}, T.~L., {Solanki}, S.~K., {Mart{\'{\i}}nez Pillet}, V.,
  {et~al.} 2010, \apjl, 723, L169

\bibitem[{{Romano} {et~al.}(2012){Romano}, {Berrilli}, {Criscuoli}, {Del Moro},
  {Ermolli}, {Giorgi}, {Viticchi{\'e}}, \& {Zuccarello}}]{2012SoPh..280..407R}
{Romano}, P., {Berrilli}, F., {Criscuoli}, S., {et~al.} 2012, \solphys, 280,
  407

\bibitem[{{Rouppe van der Voort} {et~al.}(2005){Rouppe van der Voort},
  {Hansteen}, {Carlsson}, {Fossum}, {Marthinussen}, {van Noort}, \&
  {Berger}}]{2005A&A...435..327R}
{Rouppe van der Voort}, L.~H.~M., {Hansteen}, V.~H., {Carlsson}, M., {et~al.}
  2005, \aap, 435, 327

\bibitem[{{Ruiz Cobo} \& {del Toro Iniesta}(1992)}]{1992ApJ...398..375R}
{Ruiz Cobo}, B., \& {del Toro Iniesta}, J.~C. 1992, \apj, 398, 375

\bibitem[{{S{\'a}nchez Almeida} {et~al.}(2010){S{\'a}nchez Almeida}, {Bonet},
  {Viticchi{\'e}}, \& {Del Moro}}]{2010ApJ...715L..26S}
{S{\'a}nchez Almeida}, J., {Bonet}, J.~A., {Viticchi{\'e}}, B., \& {Del Moro},
  D. 2010, \apjl, 715, L26

\bibitem[{{Scharmer} {et~al.}(2011){Scharmer}, {Henriques}, {Kiselman}, \& {de
  la Cruz Rodr{\'{\i}}guez}}]{2011Sci...333..316S}
{Scharmer}, G.~B., {Henriques}, V.~M.~J., {Kiselman}, D., \& {de la Cruz
  Rodr{\'{\i}}guez}, J. 2011, Science, 333, 316

\bibitem[{{Scharmer} {et~al.}(2008){Scharmer}, {Narayan}, {Hillberg}, {de la
  Cruz Rodr{\'{\i}}guez}, {L{\"o}fdahl}, {Kiselman}, {S{\"u}tterlin}, {van
  Noort}, \& {Lagg}}]{2008ApJ...689L..69S}
{Scharmer}, G.~B., {Narayan}, G., {Hillberg}, T., {et~al.} 2008, \apjl, 689,
  L69

\bibitem[{{Sch\"ussler}(1984)}]{1984A&A...140..453S}
{Sch\"ussler}, M. 1984, \aap, 140, 453

\bibitem[{{Shelyag} {et~al.}(2011){Shelyag}, {Keys}, {Mathioudakis}, \&
  {Keenan}}]{2011A&A...526A...5S}
{Shelyag}, S., {Keys}, P., {Mathioudakis}, M., \& {Keenan}, F.~P. 2011, \aap,
  526, A5

\bibitem[{{Shimizu} {et~al.}(2008){Shimizu}, {Lites}, {Katsukawa}, {Ichimoto},
  {Suematsu}, {Tsuneta}, {Nagata}, {Kubo}, {Shine}, \&
  {Tarbell}}]{2008ApJ...680.1467S}
{Shimizu}, T., {Lites}, B.~W., {Katsukawa}, Y., {et~al.} 2008, \apj, 680, 1467

\bibitem[{{Sigwarth} {et~al.}(1999){Sigwarth}, {Balasubramaniam}, \&
  {Kn{\"o}lker}}]{1999ASPC..183...36S}
{Sigwarth}, M., {Balasubramaniam}, K., \& {Kn{\"o}lker}, M. 1999, in
  Astronomical Society of the Pacific Conference Series, Vol. 183, High
  Resolution Solar Physics: Theory, Observations, and Techniques, ed. T.~R.
  {Rimmele}, K.~S. {Balasubramaniam}, \& R.~R. {Radick}, 36

\bibitem[{{Solanki} {et~al.}(2013){Solanki}, {Krivova}, \&
  {Haigh}}]{2013ARA&A..51..311S}
{Solanki}, S.~K., {Krivova}, N.~A., \& {Haigh}, J.~D. 2013, \araa, 51, 311

\bibitem[{{Solanki} {et~al.}(1996){Solanki}, {Zufferey}, {Lin}, {Rueedi}, \&
  {Kuhn}}]{1996A&A...310L..33S}
{Solanki}, S.~K., {Zufferey}, D., {Lin}, H., {Rueedi}, I., \& {Kuhn}, J.~R.
  1996, \aap, 310, L33

\bibitem[{{Solanki} {et~al.}(2010){Solanki}, {Barthol}, {Danilovic}, {Feller},
  {Gandorfer}, {Hirzberger}, {Riethm{\"u}ller}, {Sch{\"u}ssler}, {Bonet},
  {Mart{\'{\i}}nez Pillet}, {del Toro Iniesta}, {Domingo}, {Palacios},
  {Kn{\"o}lker}, {Bello Gonz{\'a}lez}, {Berkefeld}, {Franz}, {Schmidt}, \&
  {Title}}]{2010ApJ...723L.127S}
{Solanki}, S.~K., {Barthol}, P., {Danilovic}, S., {et~al.} 2010, \apjl, 723,
  L127

\bibitem[{{Spruit}(1979)}]{1979SoPh...61..363S}
{Spruit}, H.~C. 1979, \solphys, 61, 363

\bibitem[{{Steiner}(1999)}]{1999ASPC..184...38S}
{Steiner}, O. 1999, in Astronomical Society of the Pacific Conference Series,
  Vol. 184, Third Advances in Solar Physics Euroconference: Magnetic Fields and
  Oscillations, ed. B.~{Schmieder}, A.~{Hofmann}, \& J.~{Staude}, 38--54

\bibitem[{{Steiner} {et~al.}(1998){Steiner}, {Grossmann-Doerth}, {Kn\"olker},
  \& {Sch\"ussler}}]{1998ApJ...495..468S}
{Steiner}, O., {Grossmann-Doerth}, U., {Kn\"olker}, M., \& {Sch\"ussler}, M.
  1998, \apj, 495, 468

\bibitem[{{Takeuchi}(1999)}]{1999ApJ...522..518T}
{Takeuchi}, A. 1999, \apj, 522, 518

\bibitem[{{Title} {et~al.}(1989){Title}, {Tarbell}, {Topka}, {Ferguson},
  {Shine}, \& {SOUP Team}}]{1989ApJ...336..475T}
{Title}, A.~M., {Tarbell}, T.~D., {Topka}, K.~P., {et~al.} 1989, \apj, 336, 475

\bibitem[{{Utz} {et~al.}(2009){Utz}, {Hanslmeier}, {M{\"o}stl}, {Muller},
  {Veronig}, \& {Muthsam}}]{2009A&A...498..289U}
{Utz}, D., {Hanslmeier}, A., {M{\"o}stl}, C., {et~al.} 2009, \aap, 498, 289

\bibitem[{{Utz} {et~al.}(2010){Utz}, {Hanslmeier}, {Muller}, {Veronig},
  {Ryb{\'a}k}, \& {Muthsam}}]{2010A&A...511A..39U}
{Utz}, D., {Hanslmeier}, A., {Muller}, R., {et~al.} 2010, \aap, 511, A39+

\bibitem[{{Utz} {et~al.}(2013{\natexlab{a}}){Utz}, {Hanslmeier}, {Veronig},
  {K{\"u}hner}, {Muller}, {Jur{\v c}{\'a}k}, \&
  {Lemmerer}}]{2013SoPh..tmp....7U}
{Utz}, D., {Hanslmeier}, A., {Veronig}, A., {et~al.} 2013{\natexlab{a}},
  \solphys, 7

\bibitem[{{Utz} {et~al.}(2013{\natexlab{b}}){Utz}, {Jur{\v c}{\'a}k},
  {Hanslmeier}, {Muller}, {Veronig}, \& {K{\"u}hner}}]{2013A&A...554A..65U}
{Utz}, D., {Jur{\v c}{\'a}k}, J., {Hanslmeier}, A., {et~al.}
  2013{\natexlab{b}}, \aap, 554, A65

\bibitem[{{Utz} {et~al.}(2013{\natexlab{c}}){Utz}, {Jur{\v c}{\'a}k},
  {Bellot-Rubio}, {del Toro Iniesta}, {Thonhofer}, {Hanslmeier}, {Veronig},
  {Muller}, \& {Lemmerer}}]{2013CEAB...37..459U}
{Utz}, D., {Jur{\v c}{\'a}k}, J., {Bellot-Rubio}, L., {et~al.}
  2013{\natexlab{c}}, Central European Astrophysical Bulletin, 37, 459

\bibitem[{{van Noort}(2012)}]{2012A&A...548A...5V}
{van Noort}, M. 2012, \aap, 548, A5

\bibitem[{{Venkatakrishnan}(1986)}]{1986Natur.322..156V}
{Venkatakrishnan}, P. 1986, \nat, 322, 156

\bibitem[{{Viticchi{\'e}} {et~al.}(2010){Viticchi{\'e}}, {Del Moro},
  {Criscuoli}, \& {Berrilli}}]{2010ApJ...723..787V}
{Viticchi{\'e}}, B., {Del Moro}, D., {Criscuoli}, S., \& {Berrilli}, F. 2010,
  \apj, 723, 787

\bibitem[{{V{\"o}gler} {et~al.}(2004){V{\"o}gler}, {Bruls}, \&
  {Sch{\"u}ssler}}]{2004A&A...421..741V}
{V{\"o}gler}, A., {Bruls}, J.~H.~M.~J., \& {Sch{\"u}ssler}, M. 2004, \aap, 421,
  741

\bibitem[{{Walsh} \& {Ireland}(2003)}]{2003A&ARv..12....1W}
{Walsh}, R.~W., \& {Ireland}, J. 2003, \aapr, 12, 1

\bibitem[{{Wedemeyer} {et~al.}(2013){Wedemeyer}, {Scullion}, {Steiner}, {de la
  Cruz Rodriguez}, \& {Rouppe van der Voort}}]{2013JPhCS.440a2005W}
{Wedemeyer}, S., {Scullion}, E., {Steiner}, O., {de la Cruz Rodriguez}, J., \&
  {Rouppe van der Voort}, L.~H.~M. 2013, Journal of Physics Conference Series,
  440, 012005

\bibitem[{{Wedemeyer-B{\"o}hm} {et~al.}(2012){Wedemeyer-B{\"o}hm}, {Scullion},
  {Steiner}, {Rouppe van der Voort}, {de La Cruz Rodriguez}, {Fedun}, \&
  {Erd{\'e}lyi}}]{2012Natur.486..505W}
{Wedemeyer-B{\"o}hm}, S., {Scullion}, E., {Steiner}, O., {et~al.} 2012, \nat,
  486, 505

\bibitem[{{Wiegelmann} {et~al.}(2010){Wiegelmann}, {Solanki}, {Borrero},
  {Mart{\'{\i}}nez Pillet}, {del Toro Iniesta}, {Domingo}, {Bonet}, {Barthol},
  {Gandorfer}, {Kn{\"o}lker}, {Schmidt}, \& {Title}}]{2010ApJ...723L.185W}
{Wiegelmann}, T., {Solanki}, S.~K., {Borrero}, J.~M., {et~al.} 2010, \apjl,
  723, L185

\bibitem[{{Wiehr} {et~al.}(2004){Wiehr}, {Bovelet}, \&
  {Hirzberger}}]{2004A&A...422L..63W}
{Wiehr}, E., {Bovelet}, B., \& {Hirzberger}, J. 2004, \aap, 422, L63

\bibitem[{{Zakharov} {et~al.}(2005){Zakharov}, {Gandorfer}, {Solanki}, \&
  {L{\"o}fdahl}}]{2005A&A...437L..43Z}
{Zakharov}, V., {Gandorfer}, A., {Solanki}, S.~K., \& {L{\"o}fdahl}, M. 2005,
  \aap, 437, L43

\end{thebibliography}



\end{document}